\documentclass[draft,12pt,reqno,a4paper]{amsart}

\usepackage[margin=3cm,footskip=1cm]{geometry}

\usepackage{amssymb} \usepackage{amsmath} \usepackage{amsfonts}
\usepackage{amsthm} \usepackage{mathtools}

\usepackage{enumerate} \usepackage[latin1]{inputenc}
\usepackage[shortlabels]{enumitem}

\usepackage{color} 
\usepackage{graphicx} \usepackage{verbatim}

\DeclareMathOperator*{\swslim}{s--w^\star--lim}

\DeclareMathOperator*{\vSigmalim}{{\Sigma}--lim}

\DeclarePairedDelimiter\abs\lvert\rvert
\DeclarePairedDelimiter\norm\lVert\rVert
\DeclarePairedDelimiter\set{\{}{\}}

\newcommand{\supp}{\operatorname{supp}}

\newcommand{\Div}{{\operatorname{div}}}

\newcommand{\curl}{{\operatorname{curl}}}

\newcommand{\Ran}{{\operatorname{Ran}}}
\newcommand{\Ker}{{\operatorname{Ker}}} 
 
 \newcommand{\cs}{{\rm the Cauchy-Schwarz inequality }}
\newcommand{\cS}{{\rm the Cauchy-Schwarz inequality}}

\newcommand{\N}{{\mathbb{N}}} 
\newcommand{\R}{{\mathbb{R}}} 
\newcommand{\C}{{\mathbb{C}}}

 \renewcommand{\c}{{\rm c}}
\newcommand{\e}{{\rm e}} 
 \renewcommand{\i}{{\rm i}}
\renewcommand{\d}{{\rm d}}

\renewcommand{\Re}{{\rm Re}\,} \renewcommand{\Im}{{\rm Im}\,}

\DeclarePairedDelimiter\inp\langle\rangle

\newcommand\parb[2][]{#1 \big ( #2#1\big )} \newcommand\parbb[2][]{#1
 \Big ( #2#1\Big )}

\newcommand{\intR}{{-\!\!\!\!\!\!\int_{\rho}\,}}

 \newcommand{\vB}{{\mathcal B}}
 \newcommand{\vD}{{\mathcal D}}
\newcommand{\vE}{{\mathcal E}} \newcommand{\vF}{{\mathcal F}}
 
 \newcommand{\vH}{{\mathcal H}}
 \newcommand{\vL}{{\mathcal L}}
 
\newcommand{\vO}{{\mathcal O}} 
 \newcommand{\vR}{{\mathcal R}}

\theoremstyle{plain}
\newtheorem{thm}{Theorem}[section]
\newtheorem{proposition}[thm]{Proposition}
\newtheorem{lemma}[thm]{Lemma} 
\theoremstyle{definition} 
 \newtheorem{example}[thm]{Example}

 \newtheorem{cond}[thm]{Condition}
 \newtheorem{remark}[thm]{Remark}
\newtheorem{remarks}[thm]{Remarks}
 \newtheorem*{remarks*}{Remarks}
\newtheorem*{remark*}{Remark}

\numberwithin{equation}{section}

\title[Stationary scattering theory for $1$-body Stark operators, I]{Stationary scattering theory 
for $1$-body Stark operators, I}

\thanks{This paper is dedicated to Professor Shmuel Agmon to acknowledge his
 great influence on spectral and scattering theory of Schr\"odinger
 operators. One of us, E. Skibsted, studied several of his works in the 1980s
and had the luck to meet him personally at University of
Virginia, Charlottesville, during
longer visits in the later part of the decade and there {enjoyed} his
warm and friendly
personality as well as first-hand access to interesting preprints he was
producing at the time.}

\author{T. Adachi}
\address[T. Adachi]{Graduate School of Human and Environmental Studies, Kyoto University\\ 
Kyoto, Japan}
\email{adachi@math.h.kyoto-u.ac.jp}

\author{K. Itakura}
\address[K. Itakura]{Research Organization of Science and Technology, Ritsumeikan University\\ 
1-1-1 Noji-higashi, Kusatsu, Shiga, 525-8577, Japan}
\email{kitakura@gst.ritsumei.ac.jp}

\author{K. Ito}
\address[K. Ito]{Graduate School of Mathematical Sciences, The University of Tokyo\\
3-8-1 Komaba, Meguro-ku, Tokyo 153-8914, Japan}
\email{ito@ms.u-tokyo.ac.jp}

\author{E. Skibsted} \address[E. Skibsted]{Institut for Matematiske Fag\\
Aarhus Universitet\\ Ny Munkegade 8000 Aarhus C, Denmark}
\email{skibsted@math.au.dk}

\begin{document}

\noindent
{Dedicated to Professor Shmuel Agmon\\
}

\begin{abstract} 
We study the stationary scattering theory for a perturbed $1$-body Stark operator.
We prove existence and completeness of the stationary wave operators, 
construct the associated generalized Fourier transforms, 
and characterize asymptotics of the generalized eigenfunctions of minimal growths 
in terms of the stationary scattering matrix. 
A key element of our procedure is an improved
(possibly optimal) version of radiation condition bounds obtained previously,
which in turn is proved by a new scheme of proof. 
\end{abstract}

\maketitle
\allowdisplaybreaks

\section{Introduction}\label{sec:Introduction}

In this paper we develop the stationary scattering theory for 
a {perturbed} Stark Hamiltonian on the Euclidean space {
$\R^d$, $d\geq 2$. It is a quantum mechanical model of a charged
$d$-dimensional 
particle in a constant electric field subject further to some
one-body interaction $q=q(\cdot)$ with decay at  infinity (for example
an electron-nucleus interaction). For
simplicity the field  strength as well as the particle mass and charge
are all taken to $1$. We split and write the coordinates of $\R^d$ as
\begin{align*}
(x,y)\in\mathbb R\times \mathbb R^{d-1}=\mathbb R^d,\quad x=x_1,\ y=(y_2,\dots,y_d),
\end{align*}
 and assume that    the constant electric field points in the positive
$x$-direction and  that the  perturbation $q=q(x,y)$
satisfies  Condition~\ref{cond:one-body-starkPot}, to be  stated later.
Whence the Hamiltonian  reads
\begin{align*}
H=H_0+q;\quad H_0=\tfrac12p^2-x,\ \ p=-\i \nabla.
\end{align*} 
 Under Condition~\ref{cond:one-body-starkPot}, in particular  $H$ is self-adjoint on the Hilbert space $\vH=L^2(\R^d)$.}

We formulate and verify a stationary version of the scattering theory by using spatial asymptotics 
of time-independent quantities, such as the resolvent and generalized eigenfunctions. 
The main results include existence and completeness of the stationary wave operators.
They are applied to construction of the generalized Fourier transforms,
which diagonalize $H$, and to characterization of asymptotics of
generalized eigenfunctions. Our analysis is roughly in the spirit of
\cite{AH, GY, Sa, Sk, IS2} for usual Schr\"odinger operators; see also
\cite{II, Is1, Is2, ACH} for related results.

Our arguments are heavily dependent on certain strong radiation condition
bounds, 
 given in 
Proposition~\ref{prop:phase-space-radiation}.
These bounds have interest of their own, 
and the proofs actually constitute, compared to the literature, the
 most original part of the paper. (See \cite {AIIS1} for certain weaker radiation condition
bounds.)
 Our scheme of proof consists of exploiting a positive
 commutator argument based on commutation with a second order operator 
generated from a `good' approximate solution to the eikonal equation. 
Here `good' refers to an
 approximate solution with {`appropriate'} convexity properties. Once such a
 function is found, 
 our scheme could work for other models as well.
 An example is the $1$-body Schr\"odinger operator with a decaying long-range
 potential \cite{IS3}, which was previously studied by time-dependent methods in \cite{HS}.
Although our arguments are long, 
the techniques are quite elementary, providing presumably optimal results.
The improved results have {natural} analogues in classical mechanics, as demonstrated 
 in Subsection~\ref{subsec:200920}. 

The class of generalized
 eigenfunctions we consider is the class of `minimal' ones, which
 we show is parametrized by $\Sigma=L^2(\R_\zeta^{d-1})$. The
 components of this vector $\zeta$ play the role of
 {`asymptotic orthogonal momenta'} as demonstrated in \cite{AIIS3}. Whence the parametrization obtained of
 the class of minimal generalized
 eigenfunctions has
 a natural physical interpretation, though not to be discussed in the present paper.

{In Section   \ref{sec:settings-results} we introduce
  our  basic assumption Condition~\ref{cond:one-body-starkPot} on
  the perturbation  $q$ and 
various notation, in particular a regularized version of parabolic
variables and  various Besov spaces both  well adapted to our
problems. Our main  results,  as well as our new  radiation condition
bounds,  are all  presented in this section. In  Section
\ref{subsec:Phase space localization, classical mechanics} we study
elementary properties of  the   regularized  parabolic
variables, introduce certain phase functions and show how these
quantities lead to strong classical radiation condition
bounds,
stated in terms of certain `gamma observables'.   In Section
\ref{sec:reduct-radi-cond-1} we introduce stationary wave operators
and develop the stationary scattering theory using the  strong quantum
radiation condition
bounds of Proposition
\ref{prop:phase-space-radiation}, whose proof  finally is
given in 
Section \ref{sec:proof-prop-refprs}.}

\section{Settings and results}\label{sec:settings-results}

\subsection{Settings}
We first introduce parabolic coordinates and a comparison system. 
Throughout the paper we often work with these coordinates
as well as the standard Euclidean coordinates. 
The parabolic coordinates are known to be useful in the study of 
the Stark Hamiltonian since long ago, see e.g.\ \cite{Ti}.
We shall actually use a slightly modified version $(f,g)=(f,g_2,\dots,g_d)$ 
similar to and compensating the one used in \cite{AIIS1}.
Our comparison system is given by WKB-approximations 
of solutions to the eigenequation with purely outgoing/incoming asymptotics. 

\subsubsection{Parabolic coordinates}

Let $\breve f \in C^\infty(\R)$ be a convex function such
that $\breve f(t)=1$ for $t\leq 1/2$ and $\breve f (t)=t$ for $t\geq 2$.
 We define $f\in C^\infty(\R^d)$ as 
\begin{subequations}
 \begin{align}
f(x,y)=\breve f (r+x)^{1/2};\quad
r=(x^2+y^2)^{1/2}.
\label{eq:par1}
\end{align}
The other `parabolic variables' are defined as 
\begin{align}\label{eq:par2}
 g=f^{-1}y,\quad \text{or }\ g_i=f^{-1}y_i\ \text{ for }i=2,\dots,d.
\end{align}
\end{subequations} 
The above $f$ may be viewed as is a regularization of the original parabolic variable 
$(r+x)^{1/2}$, which has singularities along the negative $x$-axis. 
Elementary useful formulas for the parabolic coordinate change
are gathered in Subsection~\ref{subsec:200814}.

Let us present a condition on $q$ in terms of the parabolic coordinates.

\begin{cond}\label{cond:one-body-starkPot} {
The  perturbation $q$ splits into two real-valued measurable functions
as 
$q=q_1+q_2$ such that:} 
\begin{enumerate}
\item\label{item:1}
$q_1\in C^1(\mathbb R^d)$, and 
there exist $\delta\in(0,1]$ and $C>0$ such that for any $|\alpha|\le 1$
 \begin{align}\label{eq:cond20}
 |\partial_{(x,y)}^\alpha q_1|\le Cf^{-1-\delta-2\abs{\alpha}}.
 \end{align} 
\item\label{item:2}
$\mathop{\mathrm{supp}}q_2$ is compact, and $q_2(-\Delta+1)^{-1}$ is a compact operator on $\mathcal H$.
\end{enumerate}
In addition, a function $\phi\in L^2_{\mathrm{loc}}(\mathbb R^d)$ has to be 
identically zero on $\mathbb R^d$ if it satisfies 
\begin{enumerate}[i)]
\item
$(H-\lambda)\phi=0$ for some $\lambda\in\mathbb R$ in the distributional sense,
\item
$\phi=0$ on a non-empty open subset of $\mathbb R^d$. 
\end{enumerate}
\end{cond}
\begin{remarks*}
\begin{enumerate}[1)]
\item
The {latter part of Condition
  \ref{cond:one-body-starkPot} }is known as the \emph{unique continuation property},
which holds true for a fairly large class of perturbations. 
We shall not discuss it in this paper. 
\item 
Condition~\ref{cond:one-body-starkPot} is clearly sufficient for the conditions of \cite{AIIS1}, and
we will actually quote some of the results from it. 
In particular \eqref{item:1} and \eqref{item:2} imply that $q$
is $H_0$-compact, cf.\ \cite[Proposition 5.3]{AIIS1}. Whence $H$ is self-adjoint on $\vD(H_0)$.
However the radiation condition bounds \cite[Corollary~2.4]{AIIS1}
are not {strong enough} for our applications. 
Instead we will present an improved version,
Proposition~\ref{prop:phase-space-radiation}, which
 is new even for $q=0$. 
\item The bound \eqref{eq:cond20} with only $|\alpha|=0$ required is
 sufficient for the conditions of \cite{AIIS1} and in fact it suffices
 to impose this weaker condition for all of the results
 presented in Subsection \ref{subsec:200825} since those can be
 alternatively derived by a perturbative argument using the second
 resolvent equation (and strong radiation condition bounds for $q=0$). We omit the details. However for the second goal of our paper, to
 prove the strong radiation condition bounds 
 of Subsection \ref{subsec:Phase space localization, quantum mechanics}, we do need
 \eqref{eq:cond20} with $|\alpha|\leq 1$. 
\end{enumerate}
\end{remarks*}

\subsubsection{Besov spaces}
Let $\vB=\vB(f)$, $\mathcal B^*=\mathcal B^*(f)$ and $\vB^*_0=\vB^*_0(f)$ be 
the \emph{Besov spaces} with respect to the multiplication operator by the function $f$,
i.e.\ 
\begin{align*}
\begin{split}
\vB&=
\Bigl\{\psi\in L^2_{\mathrm{loc}}(\mathbb R^d)\,\Big|\, 
\sum_{n\in\N_0}2^{n/2}\|F_n\psi\|_{L^2}<\infty\Bigr\},
\\
\mathcal B^*&=
\Bigl\{\psi\in L^2_{\mathrm{loc}}(\mathbb R^d)\,\Big|\, 
\sup_{n\in\mathbb N_0}2^{-n/2}\|F_n\psi\|_{L^2}<\infty\Bigr\}
,
\\
\mathcal B^*_0
&=
\Bigl\{\psi\in \mathcal B^*\,\Big|\, \lim_{n\to\infty}2^{-n/2}\|F_n\psi\|_{L^2}=0\Bigr\}
,
\end{split}
\end{align*}
where 
$F_n=1_{\{2^n\le f<2^{n+1}\}}$ is the  characteristic function of 
the set specified by $2^n\le f<2^{n+1}$ for $n\in \mathbb N_0=\mathbb N\cup\{0\}$.
These are Banach spaces with respect to the norms
\begin{align*}
\|\psi\|_\vB
&=\sum_{n\in\N_0}2^{n/2}\|F_n\psi\|_{L^2},
\quad
\|\psi\|_{\mathcal B^*}
=\|\psi\|_{\mathcal B^*_0}
=\sup_{n\in\mathbb N_0}2^{-n/2}\|F_n\psi\|_{L^2},
\end{align*}
respectively. 
It is well known, cf. the seminal work \cite{AH}, that the inclusions
\begin{align*}
L^2_s \subsetneq \vB \subsetneq L^2_{1/2} \subsetneq L^2 
\subsetneq L^2_{-1/2} \subsetneq \vB^*_0\subsetneq \vB^* \subsetneq L^2_{-s}
;\quad 
 L^2_s=f^{-s}L^2(\mathbb R^d),
\end{align*}
hold for any $s>1/2$.
The operator $H$ does not have a nonzero generalized eigenfunction in $\mathcal B^*_0$
\cite{AIIS1},
but it does in $\mathcal B^*$ and the larger weighted $L^2$ spaces, as we will see. 
Therefore $\mathcal B^*$ is the natural space for minimal generalized eigenfunctions of $H$.

\subsubsection{WKB-approximation}

Let
\begin{align}\label{eq:sigma}
  \Sigma=L^2\parb{\R^{d-1}}=L^2\parb{\R^{d-1}_g;\mathrm dg}
\end{align}
be the space of the $L^2$-functions only of $g$.
{Throughout the paper for any $\kappa>0 $ we let 
$\chi(\cdot<\kappa)$ be a smooth real-valued function $\chi$ on $\R$
such that
\begin{subequations}
\begin{align}
0\leq \chi\leq 1,\quad \supp \chi \subseteq (-\infty,\kappa),\quad 
\chi(t)=1\ \,\text{for }\, t\leq 3\kappa/4.
\label{eq:20100311}
\end{align}
In addition we introduce a smooth cut-off function that complements $\chi(\cdot<\kappa)$ by 
\begin{align}
\chi^\perp(\cdot<\kappa)=1-\chi(\cdot<\kappa).
\label{eq:20100312}
\end{align} 
\end{subequations}
For any $\lambda\in\mathbb R$ and $\xi\in \Sigma$ we set 
\begin{align}\label{eq:gen1B}
\begin{split}
\phi_\lambda^\pm[\xi](f,g)=
\omega_\pm^{-1}\chi^{\perp}(f<2) J(f,g)^{1/2}\,\e^{\pm\i\theta_\lambda(f)}
\,\xi(\pm g),
\end{split}
\end{align} 
where
\begin{align}\label{eq:Jac1}
\omega_\pm=(2\pi)^{1/2}\e^{\pm\i\pi d/4},
\quad 
J(f,g)
&=f^{2-d}(f^2+g^2)^{-1}
,\quad
\theta_\lambda(f)
=\tfrac13f^3+\lambda f.
\end{align}
}We shall use \eqref{eq:gen1B} as a comparison system for our stationary scattering theory. 
Here we {assert, to be elaborated on below,} that the functions $\phi^\pm_\lambda[\xi]$ 
belong to $\mathcal B^*$ and constitute (zeroth order) WKB-approximations of 
solutions to the eigenequation with purely outgoing/incoming radiation conditions,
respectively.

The former assertion follows from the fact that
 $J$ is the Jacobian associated with the 
coordinate change between $(x,y)$ and $(f,g)$: 
For any $\psi\in \mathcal H$ supported in $r+x> 2$
we can compute, using formulas from Subsection~\ref{subsec:200814}, 
\begin{align*}
\int|\psi(x,y)|^2\,\mathrm dx\mathrm dy
=
\int|\psi(f,g)|^2 J(f,g)^{-1}\,\mathrm df\mathrm dg.
\end{align*}
On the other hand, the \textit{phase function} $\theta_\lambda$ is 
an approximate solution to the eikonal equation
\begin{subequations}
\begin{align}
\tfrac12|\nabla \theta|^2-x=\lambda
\label{eq:eik}
\end{align}
in the sense that it satisfies (by formulas from Lemma \ref{lem:200821}) for $g$ bounded
\begin{align}\label{eq:eik2}
\tfrac12|\nabla \theta_\lambda|^2-x
&
=
\lambda
+\mathcal O(f^{-2}).
\end{align}
\end{subequations}
We can actually find an exact solution $\theta=\theta^{\mathrm{ex}}_\lambda$ to \eqref{eq:eik}, 
see Subsection~\ref{subsec:200819}. 
However, the exact solution $\theta^{\mathrm{ex}}_\lambda$ is not
globally defined on $\mathbb R^d$, and 
this could suggest that a regularized version like 
$\theta_\lambda$ is more useful for applications in quantum
mechanics. We are not
going to use the exact solution at all, and in fact we will demonstrate
certain strong radiation condition bounds in terms of observables
constructed entirely from the function $\theta_\lambda$.

Next, let us introduce the first order differential operators 
\begin{align}
\partial_f=2r(\nabla f)\cdot\nabla
,\ \ 
p_f=-\mathrm i\partial_f
,\ \ 
B
=\mathop{\mathrm{Re}}p_f
=2r(\nabla f)\cdot p-\mathrm i\mathop{\mathrm{div}}(r\nabla f).
\label{eq:2008075b}
\end{align} (Note that the coefficients are smooth even at $r=0$ by
our construction of $f$.)
Although $\partial_f$ does not coincide with the {coordinate derivative
$\tfrac{\partial}{\partial f}$} 
for $r+x\leq 2$, 
it does for $r+x> 2$ with expressions
\begin{align*}
\partial_f=f\partial_x+g\partial_y,
\ \ 
B=p_f-\tfrac{\mathrm i}2fr^{-1}-\tfrac{\mathrm i(d-2)}2f^{-1}
,
\end{align*}
see Lemma \ref{lem:200821}. 
Then for $r+x> 2$ we obtain $BJ^{1/2}=0$, 
and hence the \emph{outgoing/incoming radiation conditions} 
\begin{align}
\bigl(B\mp (\partial_f\theta_\lambda)\bigr) \phi_\lambda^\pm[\xi]=0,
\label{eq:200816}
\end{align}
respectively. 
Here we have used that $\partial_fg=0$ for $r+x> 2$. 
We remark that $B$ coincides with the generator of the semigroup of unitary $f$-translations on $\mathcal H$
for $r+x> 2$, although we will not use this property in the paper.

Finally we verify that $\phi_\lambda^\pm[\xi]$ for $\xi\in C^\infty_{\mathrm c}(\R^{d-1})$ 
is an \emph{approximate} generalized eigenfunction in the sense that 
\begin{align}
\psi^\pm_\lambda[\xi]:=(H-\lambda)\phi_\lambda^\pm[\xi] \in \vB. 
\label{eq:200815}
\end{align}
With {the repeated index summation convention} assumed, let 
\begin{align}
L=p_i\ell_{ij}p_j,\text{ where }\ell_{ij}=\delta_{ij}-2r(\partial_if)(\partial_jf)\ \ \text{for }i,j=1,\dots,d.
\label{eq:200818}
\end{align}
Using \eqref{eq:eik2} we can then decompose for $g$ bounded and $r+x> 2$
\begin{align}
H-\lambda
&
=
\tfrac14\bigl(B\pm (\partial_f\theta_\lambda)\bigr)r^{-1}\bigl(B\mp (\partial_f\theta_\lambda)\bigr)
+\tfrac12L
+\mathcal O(f^{-1-\delta})
.
\label{eq:2009212345}
\end{align}
{We note that $L$ does not act on the variable $f$ due to 
the fact that $|\nabla f|^2=\tfrac12r^{-1}$, cf.\ Lemma \ref{lem:200821}.
Then along with \eqref{eq:200816} and Lemma \ref{lem:20092016} we certainly obtain \eqref{eq:200815}}.

\subsection{Main results}\label{subsec:200825}

We present the main results of the paper. 
The corresponding results for decaying potentials are studied in
several papers, see for example \cite{Sk, IS2} and the references given there.

\subsubsection{Stationary scattering theory} \label{subsec:Wave operators}

We first discuss the stationary scattering theory. 
We provide existence and completeness of the stationary wave operators, 
and then introduce the stationary scattering matrix. 

{The resolvent at any  $z\in \C\setminus \R$ is denoted by $R(z)=(H-z)^{-1}$. By \cite{AIIS1} for any $\lambda\in\mathbb R$ the following two limiting resolvents exist, 
\begin{align*}
R(\lambda\pm\mathrm i0)=\swslim_{\mu \downarrow 0,\,z=\lambda\pm
 \mathrm i \mu}R(z)\in
 \mathcal L(\mathcal B,\mathcal B^*), \text{ respectively}.
\end{align*} 
Here the limits are given in the sense of `strong weak-star'
limits, and in the strong weak-star topology they are continuous in $\lambda$. 
The word `respectively' is typically omitted in the following.
We are going to construct the stationary wave operators by 
comparing $R(\lambda\pm\mathrm i0)$ with the WKB-approximation from \eqref{eq:gen1B}.
For any $\psi\in\mathcal B$ we introduce
\begin{align*}
 \vR^\pm_\lambda\psi(f,g)
=\omega_\pm
 J(f,g)^{-1/2}\e^{\mp\i\theta_\lambda(f)}\parb{R(\lambda\pm\mathrm
 i0)\psi}(f, g)
\end{align*}
 and might expect, recalling the notation \eqref{eq:sigma},} that there exist limits
\begin{align}\label{eq:f_limit} 
\vSigmalim_{f\to\infty}\vR^\pm_\lambda\psi(f,\cdot)
:=\lim_{f\to \infty}\vR^\pm_\lambda\psi(f,\cdot) \text{ in } \Sigma.
\end{align} 
In fact we do take the limits \eqref{eq:f_limit}, however in general only in the averaged sense. 
For any vector-valued function $\xi$ of $f$ let us use the notation 
$$\intR \xi(f)\,\d f=\rho^{-1}\int_{\rho}^{2\rho} \xi(f)
\,\d f,
\quad 
\rho>0.$$

\begin{thm}\label{thm:dist-four-transf} 
\begin{enumerate}
\item
For any $\lambda\in\R$ and $\psi\in \vB$ 
{the following averaged limits exist,}
\begin{align}\label{eq:extbF2}
\begin{split} 
\vF^\pm(\lambda)\psi 
:= \pm(2\pi\mathrm i)^{-1}\,\vSigmalim_{\rho\to
 \infty} {-\!\!\!\!\!\!\int_\rho} \vR^{\pm}_{\lambda}\psi(f,\pm\cdot)\,\d f.
\end{split}
\end{align}

\item
The mappings $\mathbb R\times \mathcal B\ni
(\lambda,\psi)\mapsto \mathcal F^\pm (\lambda)\psi\in\Sigma$ 
are continuous.

\item
For any $\lambda\in\mathbb R$ the operators $\vF^\pm(\lambda)\colon \vB\to \Sigma$ 
are surjective.
 
\item\label{item:delta}
For any $\lambda\in\mathbb R$ and $\psi\in \vB$
\begin{align}
 \label{eq:fund}
 \|\vF^\pm(\lambda)\psi\|^2=\inp{\psi, \delta(H-\lambda)\psi},
\end{align} 
where $\delta(H-\lambda)=\pi^{-1}\mathop{\mathrm{Im}}R(\lambda+\i 0)$.
\end{enumerate}

\end{thm}

The operators $\vF^\pm(\lambda)$ are called the \emph{stationary wave operators},
and their surjectivity is the \emph{stationary completeness}. 
In addition, the adjoints $\vF^\pm(\lambda)^*\in \vL(\Sigma, \vB^*)$ 
are called the \emph{stationary wave matrices}. 
Using the properties from Theorem~\ref{thm:dist-four-transf}, 
we can further define the \emph{stationary scattering matrix} $S(\lambda)\in \mathcal L(\Sigma)$
for $\lambda\in\mathbb R$ as the unique unitary operator obeying
\begin{align}\label{eq:scattering_matrix}
 {\vF^+(\lambda)\psi}=S(\lambda){\vF^-(\lambda)\psi}\ \ \text{for any }\psi\in\mathcal B.
\end{align} 
By construction the mapping 
{$\R \ni\lambda\mapsto S(\lambda) \in \vL(\Sigma)$} is strongly continuous.

\begin{example}\label{example:stat-scatt-theory}
 For $q=0$ the operators $\vF^+(\lambda)=\vF^-(\lambda)$,
and hence $S(\lambda)=I$. This fact
 follows from the Airy-Fourier transform
 and the stationary phase
 method, and it is part of a more detailed study of the scattering
 matrix in \cite{AIIS3}. 
\end{example}

\subsubsection{Generalized Fourier transforms}
Using the stationary wave operators we can 
construct the \emph{generalized Fourier transforms} that diagonalize $H$. 
 Introduce 
\begin{align*}
{\mathcal F^\pm=\int_{\mathbb R}^\oplus \mathcal F^\pm(\lambda)\,\mathrm d\lambda,\quad 
 \widetilde \vH =L^2(\R, \d \lambda;\Sigma),}
\end{align*} 
and let $M_\lambda$ denote the multiplication operator by $\lambda$ on $\widetilde\vH$.

\begin{thm}\label{thm:unit-equiv}
The operators $\mathcal F^\pm$ extend as 
unitary operators $\vF^\pm\colon \mathcal H\to \widetilde{\mathcal H}$,
and further satisfy
$$\vF^\pm H=M_\lambda \vF^\pm.$$
In particular, $H$ and $M_\lambda$ are unitarily equivalent. 
\end{thm}

\subsubsection{Generalized eigenfunctions of minimal growth} \label{subsec:Wave operatorsbb}

We let 
\begin{equation*}
 \vE_\lambda=\{\phi\in \vB^*\,|\, (H-\lambda)\phi=0\}, 
\end{equation*}
and call {the  elements of this space }\emph{minimal generalized eigenfunctions}. 
They are minimal in the sense that there does not exist a generalized eigenfunction in 
the slightly smaller space $\mathcal B^*_0$, \cite{AIIS1}. 
The following result shows $\mathcal E_\lambda\neq\{0\}$ and 
characterizes the asymptotics of {the
 eigenfunctions in this space} in terms of $S(\lambda)$. 
We remark that the existence of nonzero elements in $\mathcal E_\lambda$ is non-trivial even if $q=0$. 
For instance, the Airy function $\phi(x,y)=\mathop{\mathrm{Ai}}(x)$ 
 does not belong to $\mathcal B^*$. 

\begin{thm}
 \label{thm:char-gener-eigenf-1}
\begin{subequations}
 \begin{enumerate}[(1)]
 \item\label{item:14.5.13.5.40} For any one of $\xi_\pm \in \Sigma$ or $\phi\in \vE_\lambda$ the
 two other quantities in $\{\xi_-,\xi_+, \phi\}$ uniquely exist
 such that
 \begin{align}\label{eq:gen1}
 \phi -\phi_\lambda^+[\xi_+]-\phi_\lambda^-[\xi_-]\in
 \vB_0^*.
 \end{align}

 \item \label{item:14.5.13.5.41} The correspondences in \eqref{eq:gen1} are given by the
 formulas
 \begin{align}\label{eq:aEigenfw}
 \phi&= \vF^\pm(\lambda)^*\xi_\pm,\qquad \xi_+=S(\lambda)\xi_-,\\
\xi_\pm&= {\omega_\pm\,\vSigmalim_{\rho\to \infty}
-\!\!\!\!\!\!\int_\rho \,\bigl (J^{-1/2}\,\e^{\mp\i\theta_\lambda}
\bigl(\tfrac12\pm \tfrac12(f\sqrt{2r})^{-1}p_f\bigr)\phi\bigr)(f,\pm\cdot)\,\d f.}\label{eq:aEigenfwB}
 \end{align}

 \item\label{item:14.5.13.5.42} The wave matrices
 $\vF^\pm(\lambda)^*\colon\Sigma\to \vE_\lambda\,(\subseteq \vB^*)$
 are topological linear isomorphisms.
 In addition, for any $\xi_\pm\in\Sigma$ and $\phi\in\mathcal E_\lambda$ satisfying \eqref{eq:gen1}
 \begin{align}\label{eq:aEigenf2w}
 \|\xi_\pm\|_{\Sigma}=\pi^{1/2}\lim_{n \to \infty}2^{-n/2}\norm{F_n\phi}_{L^2},
 \end{align}
where $F_n$ is the characteristic function $F_n=1_{\{2^n\le f<2^{n+1}\}}$. 

 \item\label{item:14.5.14.4.17} 
The operator
$\delta(H-\lambda)=\pi^{-1}\mathop{\mathrm{Im}}R(\lambda+\i
0):\,\vB\to \vE_\lambda$ is surjective.
 \end{enumerate}
 \end{subequations}
\end{thm}

We finally provide some supplementary formulas involving the vector 
$$\psi_\lambda^\pm[\xi]
=(H-\lambda)\phi_\lambda^\pm[\xi] \in \vB\ \ \text{with }\xi\in C^\infty_{\mathrm c}(\mathbb R^{d-1}), $$
cf.\ \eqref{eq:200815}. 
We refer to \cite{AIIS3} for other, somewhat similar formulas.

\begin{proposition}\label{cor:scatt-matr-gener} 
 \begin{subequations}
For any $\xi \in C^\infty_{\mathrm c}(\R^{d-1})$
\begin{align}\label{eq:rep1-} 
 \vF^\pm(\lambda)^* \xi
= &\phi_\lambda^\pm[\xi]-R(\lambda\mp\i 0)\psi_\lambda^\pm[\xi]
=2\pi\i \delta(H-\lambda)\psi_\lambda^\pm[\xi],\\
\label{eq:rep3-}
 \xi= &\pm 2\pi\i \vF^\pm(\lambda)\psi_\lambda^\pm[\xi],\\
 \label{eq:formS1}
S(\lambda)\xi=&-2\pi\i\vF^+(\lambda)\psi_\lambda^-[\xi]=- \vSigmalim_{\rho\to
 \infty} \intR \vR_{\lambda}^{+}\psi_\lambda^-[\xi](f,\cdot)\,\d f.
 \end{align}
For any $\xi, {\xi'} \in C^\infty_{\mathrm c}(\R^{d-1})$ 
\begin{align}\label{eq:formS2}
 \begin{split}
 \tfrac 1{2\pi \i}\inp{{\xi'},S(\lambda)\xi}=\inp{\psi_\lambda^+[{\xi'}],R(\lambda+
 \i
 0)\psi_\lambda^-[\xi]}-\inp{\phi_\lambda^+[{\xi'}],\psi_\lambda^-[\xi]}.
\end{split}
\end{align}
\end{subequations} 
 \end{proposition}

\subsection{Key estimates: Radiation condition bounds}\label{subsec:Phase space localization, quantum mechanics}

The proofs of our main results are heavily dependent on 
the following \textit{radiation condition bounds}
as well as results from \cite{AIIS1}.
Such bounds were considered also in \cite{AIIS1}, 
but here we provide a considerably improved version. 
Once this strong version is at our disposal,
the construction of stationary wave operators is almost trivial 
along the lines of \cite{HS, Sk,IS2}. 
The radiation condition bounds themselves are of independent interest,
and may be considered as one of our main results. 
In fact, their proof requires an original idea
and occupies a substantial part of the paper. 

We introduce for any $\lambda\in\mathbb R$ `gamma observables' as 
\begin{subequations}
\begin{align}\label{eq:QM1}
\gamma(\lambda)
=p\mp\nabla \theta_\lambda
,\quad
\tilde\gamma=f^{-1}g
\end{align} 
and the derived observable
\begin{align}
 \begin{split}
\gamma_\|(\lambda)
&=\mathop{\mathrm{Re}}\bigl(2r(\nabla f)\cdot\gamma(\lambda)\bigr)
=2r(\nabla f)\cdot\gamma(\lambda)- \mathrm
i\mathop{\mathrm{div}}(r\nabla f)
=B-\partial_f\theta_\lambda,
\label{eq:200820} 
 \end{split}
\end{align} 
\end{subequations} 
cf.\ \eqref{eq:2008075b} and \eqref{eq:200816}.
We also refer to these quantities as \emph{radiation operators}. 
These operators are quantizations of corresponding classical observables 
 to be considered in Subsection~\ref{subsec:200920}.

\begin{proposition}\label{prop:phase-space-radiation} 
Let $I\subseteq \mathbb R$ be a compact interval, and let $s\in (0,1+\delta)$. 
Then there exists $C>0$ such that for any $i,j=1,\dots,d$ and any $\phi=R(\lambda\pm\i 0)\psi$
with $\lambda\in I$ and $\psi\in C_{\mathrm c}^\infty(\mathbb R^d)$
\begin{align}
\begin{split}
&\norm{f^s\gamma_i(\lambda)\gamma_j(\lambda)\phi}_{L^2_{-1/2}}
+{\norm{f^{(s+2)/2}\tilde\gamma^2\phi}_{\mathcal B^*}}
+\norm{f^s\gamma_\|(\lambda)\phi}_{L^2_{-1/2}}
\\&\quad \quad \quad \quad 
\le C\|f^{s+2}\psi\|_{L^2_{1/2}}.
\end{split}
\label{eq:p1}
\end{align}
 \end{proposition}
\begin{remarks}\label{rem:20103018}
\begin{enumerate}[1)]
\item\label{item:2010301809}
The action of the radiation operators in \eqref{eq:p1} is understood in the distribution sense. 
{While it is not claimed, e.g.,\ that 
$f^s\gamma_\|(\lambda)R(\lambda\pm\i \mu)\psi$
 converges to $f^s\gamma_\|(\lambda)\phi$
 in $L^2_{-1/2}$ as $\mu\downarrow 0$, 
we can show its uniform boundedness in $L^2_{-1/2}$ if we restrict to $s\in (0,1)$. 
See Remarks~\ref{rem:20:1030} and \ref{remarks:commutator-estimates} \ref{item:20103015}.}
\item\label{item:2010301810}
We can further deduce finiteness of several weighted {$\mathcal B^*$-norms}.
For example for any $i=1,\dots,d$ and
$j=2,\dots,d$ the norms
 \begin{align}
 \begin{split}
 &
\norm{f^{s/2}\gamma_i(\lambda)\phi}_{\mathcal B^*},\quad 
\norm{f^{(s+2)/4}\tilde\gamma_j\phi}_{\mathcal B^*},\quad 
\norm{f^{(3s+2)/4}\tilde\gamma_j\gamma_i(\lambda)\phi}_{\mathcal B^*}
\label{eq:p3} 
 \end{split}
\end{align} 
are all finite (with the same bounding constant). 
Formally these estimates follow by integration by
parts, the Cauchy-Schwarz inequality, \eqref{eq:p1} and the 
LAP bound of the form \cite[Theorem~2.8]{AIIS1}.
See Remark \ref{remarks:commutator-estimates} \ref{item:bProof} for
{an elaboration}.
\item
The weights are considerably better than those of \cite[Corollary~2.14]{AIIS1}. 
Let $\delta=1$ for simplicity. 
Then the exponent $\beta$ in \cite{AIIS1} has an upper bound $\beta_c=1/2$, 
but the corresponding values are $\beta_c=2$ for the first norm of \eqref{eq:p3},
and $\beta_c=4$ for the last term on the left-hand side of
\eqref{eq:p1}. We believe these improved bounds are `optimal', although we do
not prove it.
\item
In order to prove such an improved version
we invent a commutator method involving a second order operator
similarly to \cite{IS3},
whereas in \cite{AIIS1} the standard first order one 
was considered.
Our scheme consists only of elementary techniques such as 
the Leibniz rule and the Cauchy-Schwarz inequality
without microlocal analysis or advanced functional analysis.
It also has a {natural  analogue in classical mechanics 
to be demonstrated in Subsection~\ref{subsec:200920}, providing
analogous results}. 
\end{enumerate}
\end{remarks}

\section{Classical mechanics}\label{subsec:Phase space localization, classical mechanics}

In this section we first provide useful elementary formulas for the parabolic coordinates 
and then discuss the classical mechanics. 
In particular, we present an exact solution to the eikonal equation with the Stark field. 
Although we do not use the exact solution in the
 later sections on quantum mechanics, we give in Subsection
 \ref{subsec:200920} a motivating
 discussion involving the exact solution. We show how our
 approximate solution provides strong radiation condition bounds in
 classical mechanics by a differential inequality technique that in
 fact can be extended to quantum mechanics.

\subsection{Formulas for parabolic coordinates}\label{subsec:200814}

Here we gather formulas to be used many times in the sequel. 
All of them are direct consequences of the definitions \eqref{eq:par1} and \eqref{eq:par2}.
The first result concerns some global bounds.
\begin{subequations}

\begin{lemma}\label{lem:20092016}
For any $t\in\R$ and $\alpha\in\mathbb N_0^d$
there exists $C_{t,\alpha}\ge 0$ such that 
\begin{align}\label{eq:basDERa}
\abs{\partial^\alpha f^t} \leq C_{t,\alpha} f^{t-\abs{\alpha}}\langle r\rangle^{-|\alpha|/2};\quad 
\inp{r}=(1+r^2)^{1/2}.
\end{align} 
For any $i=2,\dots, d$ 
and any $\alpha\in\mathbb N_0^d$ there exists $C_{i,\alpha}\ge 0$ such that
\begin{align}
 \label{eq:gboundfb}
\abs{\partial^\alpha g_i} \leq C_{i,\alpha} f^{-\abs{\alpha}}\langle r\rangle^{1/2-|\alpha|/2}.
\end{align}

\end{lemma}
 
\end{subequations}
\begin{proof}
In this proof we let $\phi=r+x$. We first observe that 
 \begin{align*}
 2r\phi-y^2=r(r+x)+r^2+rx-y^2=(r+x)^2\ge 0,
 \end{align*}  implying  that
\begin{align}
 \label{eq:batri}
 \abs{y}\leq \sqrt{2r\phi}\le \sqrt2 f \sqrt{r}.
 \end{align}
 Clearly \eqref{eq:gboundfb} for $\alpha=0$ follows from
 \eqref{eq:batri}. For $\alpha\neq 0$ \eqref{eq:gboundfb} follows readily from
 \eqref{eq:basDERa} with $t=-1$ in combination with \eqref{eq:batri}.
 
 Thus it suffices to show \eqref{eq:basDERa}, 
and this in fact reduces to the case $t=2$ by Fa\`a di Bruno's formula
{(see for example \cite{EM})}
applied to the composition $f^t=(f^2)^{t/2}$. 
By the same formula now applied to $\breve f$ and $\phi$,
see \eqref{eq:par1}, 
it further reduces to the statement that for any $|\alpha|\ge 1$
 \begin{align}\label{eq:baaaaa}
 \abs{{\partial^\alpha \phi} }\leq C_\alpha f^{2-\abs{\alpha}}\langle
 r\rangle^{-|\alpha|/2}\ \ \text{if }\phi\ge 1/2.
 \end{align}
To prove \eqref{eq:baaaaa} for $\abs{\alpha}=1$ 
we compute 
\begin{align*}
\nabla \phi =(\phi/r,y/r),\quad 
|\nabla \phi|^2=r^{-2}((r+x)^2+y^2)=2\phi/r.
\end{align*}
Since $r\ge 1/4$ if $\phi\ge 1/2$, we obtain \eqref{eq:baaaaa} for $|\alpha|=1$. 
It is easy to see 
$$|\partial^\alpha\phi|\le C_\alpha'\langle r\rangle^{1-|\alpha|}, $$
since $\phi$ is homogeneous of degree $1$ and $r\ge 1/4$ if $\phi\ge 1/2$.
This bound implies \eqref{eq:baaaaa} for $|\alpha|\ge 2$ also, and we are done. 
 \end{proof}

On a set where $f$ is large (more precisely for $f>\sqrt 2$) we have various explicit formulas.

\begin{lemma}\label{lem:200821}
The following identities and bounds hold on the open set $r+x> 2$. 
One has 
\begin{align}\label{eq:eikalmost}
 f^2+g^2=2r,\quad f^2-g^2=2x.
\end{align} 
For any $i,j=2,\dots,d$
$$\partial_fx=f,\quad 
\partial_{g_i}x=-g_i,\quad 
\partial_fy_i=g_i,\quad 
\partial_{g_i}y_j=\delta_{ij}f,$$
so that 
$$p_f=fp_x+g_ip_{y_i}\,\parb{=fp_x+\sum_{i\geq 2} \,g_ip_{y_i}},\quad 
p_{g_i}=-g_ip_x+fp_{y_i}.$$
For any $i,j=2,\dots,d$
\begin{align*}
\begin{split}
\partial_xf
&=
\tfrac12 fr^{-1}
,\ \ 
\partial_{y_i}f
=\tfrac12g_ir^{-1}, \ \ 
\partial_xg_i=-\tfrac12 g_ir^{-1},\ \ 
\partial_{y_i}g_j=f^{-1}[\delta_{ij}-\tfrac12 r^{-1}g_ig_j],
\end{split}
\end{align*}
so that 
\begin{align*}
p_x=\tfrac12fr^{-1}p_f-\tfrac12g_jr^{-1}p_{g_j},\quad 
p_{y_i}=\tfrac12g_ir^{-1}p_f+f^{-1}(\delta_{ij}-\tfrac12r^{-1}g_ig_j)p_{g_j}.
\end{align*}
As for the second derivatives, for any $i,j=1,\dots,d$
\begin{align}
 \label{eq:seDER}
 \partial_i\partial_jf
=\tfrac12f^{-1}r^{-1}\bigl(\delta_{ij}-2r(\partial_if)(\partial_jf)- (\partial_ir)(\partial_jr)\bigr).
\end{align} 
In particular, for any $i=2,\dots,d$
\begin{align}
\begin{split}
&|\nabla f|^2=(2r)^{-1},\quad 
(\nabla f)\cdot (\nabla g_i)=0,\quad 
\\ &
(\nabla f)\cdot (\nabla r)=(2r)^{-1}f,\quad 
\Delta f
=\tfrac{d-2}2f^{-1}r^{-1},
\end{split}
\label{eq:ort}
\end{align}
and 
\begin{align}
\tfrac12fr^{-1}I
\le h:=2\mathop{\mathrm{Re}}(\nabla r\nabla f)
\le f^{-1}I.
\label{eq:20090623}
\end{align}
\end{lemma}
\begin{proof}
These are also directly computed from the definitions \eqref{eq:par1} and \eqref{eq:par2}. 
Here we verify only \eqref{eq:20090623}. 
By definition and the above formulas we have for any $i,j=1,\dots,d$
\begin{align*}
h_{ij}&=2r\partial_i\partial_j f
+(\partial_ir)(\partial_j f)
+(\partial_i f)(\partial_jr)
\\&
=
f^{-1}\delta_{ij}
-2f^{-1}r(\partial_if)(\partial_jf)
-f^{-1}(\partial_ir)(\partial_jr)
+(\partial_ir)(\partial_j f)
+(\partial_i f)(\partial_jr)
\\&
=
f^{-1}\delta_{ij}
-\tfrac14f^{-3}g^2(\partial_if^2)(\partial_jf^2)
-\tfrac14f^{-1}(\partial_ig^2)(\partial_jg^2)
\\&
=
\tfrac12 fr^{-1}\delta_{ij}
+\tfrac12 f^{-1}g^2r^{-1}\delta_{ij}
-\tfrac14f^{-3}g^2(\partial_if^2)(\partial_jf^2)
-\tfrac14f^{-1}(\partial_ig^2)(\partial_jg^2).
\end{align*} 
Here we have used the first formula of \eqref{eq:eikalmost} 
to obtain the third and fourth equalities.
Clearly $h\le f^{-1}I$ follows from the expression of the third line,
and we can deduce $\tfrac12fr^{-1}I\le h$ from that of the fourth line. Indeed, for the latter, 
it suffices to 
note that $\nabla f^2$ and $\nabla g^2$ are orthogonal to each other and that 
$$
\tfrac14f^{-3}g^2|\nabla f^2|^2
=\tfrac14f^{-1}|\nabla g^2|^2
=\tfrac12f^{-1}g^2r^{-1},
$$ yielding that $ h-\tfrac12fr^{-1}I=\tfrac12 f^{-1}g^2r^{-1}P$
for some orthogonal projection $P$.
\end{proof}

\subsection{Phase function}\label{subsec:200819}

Motivated by a stationary phase
 analysis of the Airy-Fourier transform, cf.\ \cite{AIIS3}, we
 introduce the phase function
\begin{align}
{\theta^{\mathrm{ex}}(x,y)
=\tfrac43\sqrt{x+\parb{x^2-y^2}^{1/2}}\parbb{x-\tfrac12\parb{x^2-y^2}^{1/2}}
\label{eq:20070210};\quad  x>\abs{y}.}
\end{align} 
This is an exact solution to the eikonal equation \eqref{eq:eik} with $\lambda=0$. 
Here we restrict ourselves to the zero-energy case for simplicity,
but the general case can be easily reproduced by $x$-translation 
letting $\theta^{\mathrm{ex}}_\lambda(x,y)=\theta^{\mathrm{ex}}(x+\lambda,y)$.

\begin{lemma}\label{lem:20062814}
The function $\theta^{\mathrm{ex}}$ solves the eikonal equation \eqref{eq:eik} with $\lambda=0$, i.e. 
\begin{align*}
 \tfrac 12|\nabla\theta^{\mathrm{ex}}|^2 - x =0.
\end{align*}
\end{lemma}
\begin{proof}
This is due to direction computations. 
We can compute
\begin{align*}
\begin{split}
\nabla\theta^{\mathrm{ex}} 
&= \tfrac 43 \frac{x - \tfrac 12 \parb{x^2 - y^2}^{1/2}}{2\sqrt{x+ \parb{x^2 - y^2}^{1/2}}}{}\left( 1+x\parb{x^2 - y^2}^{-1/2}, -y\parb{x^2 - y^2}^{-1/2} \right) \\
 &\quad + \tfrac 43 \sqrt{x+ \parb{x^2 - y^2}^{1/2}}\left( 1 - \tfrac x2 \parb{x^2 - y^2}^{-1/2}, \tfrac y2 \parb{x^2 - y^2}^{-1/2} \right) \\
 &= \sqrt{x+ \parb{x^2 - y^2}^{1/2}}\left( 1, y\left( x+ \parb{x^2 - y^2}^{1/2} \right)^{-1}\right),
\end{split}
\end{align*}
and hence 
\begin{align*}
|\nabla\theta^{\mathrm{ex}}|^2 
&= \left( x+ \parb{x^2 - y^2}^{1/2} \right)\left( 1+y^2\left( x+ \parb{x^2 - y^2}^{1/2} \right)^{-2} \right) 
 = 2x,
\end{align*}
which implies the assertion.
\end{proof}

Although we have an exact and explicit solution $\theta^{\mathrm{ex}}$, 
the expression is fairly complicated and is not globally defined. 
The following lemma somehow guarantees that 
the regularization $\theta_0$ from \eqref{eq:Jac1} approximates well
$\theta^{\mathrm{ex}}$. 
Let us set
\begin{align*}
\Omega 
=\{x>|y|\}\cap \{r+x> 2\}
=\bigl\{g^2/f^2<3-2\sqrt{2}\bigr\}\cap \{r+x> 2\}.
\end{align*}

\begin{lemma}
There exists an analytic function $F(z)$ in $|z|<3-2\sqrt2$ such that 
\begin{align*}
\theta^{\mathrm{ex}}=f^3\bigl(\tfrac13+(g^4/f^4)F(g^2/f^2)\bigr)
\ \ 
\text{in }\Omega.
\end{align*}
In particular, for any $\kappa>0$ and $\alpha\in\mathbb N_0^d$ there exists $C_{\alpha\kappa}>0$ such that 
\begin{align}\label{eq:comfs}
|\partial^\alpha(\theta^{\mathrm{ex}}-\theta_0)|
\le C_{\alpha\kappa} f^{-1-\max\{|\alpha|,2|\alpha|-4\}}
\ \text{ at points in } \Omega\text{ with }|g|\le \kappa.
\end{align}
\end{lemma}
\begin{proof}
We rewrite $\theta^{\mathrm{ex}}$ in terms of the parabolic coordinates $(f,g)$.
By setting $z=g^2/f^2$ we can write 
\begin{align*}
\theta^{\mathrm{ex}}
&
=
\tfrac23\sqrt{\tfrac12f^2-\tfrac12g^2+ \tfrac12\parb{f^4-6f^2g^2+g^4}^{1/2}}
\parbb{f^2-g^2-\tfrac12\parb{f^4-6f^2g^2+g^4}^{1/2}}
\\&
=
\tfrac23f^3
\sqrt{\tfrac12-\tfrac12z+ \tfrac12\parb{1-6z+z^2}^{1/2}}
\parbb{1-z-\tfrac12\parb{1-6z+z^2}^{1/2}}.
\end{align*}
The last function is analytic in $|z|<3-2\sqrt2$, and we can expand it into the Taylor series. 
We actually have 
\begin{align*}
\sqrt{\tfrac12-\tfrac12z+ \tfrac12\parb{1-6z+z^2}^{1/2}}
\parbb{1-z-\tfrac12\parb{1-6z+z^2}^{1/2}}
=\tfrac12+\tfrac34z^2+\cdots, 
\end{align*}
and this implies the former assertion. 
We obtain \eqref{eq:comfs} by using Lemma \ref{lem:20092016}.
\end{proof}

\subsection{Phase space localization}\label{subsec:200920}

Now we study the {asymptotic behavior} of classical scattering  orbits in the phase space.
The estimates below are not necessary for the later arguments, but 
are the original motivation for Proposition~\ref{prop:phase-space-radiation}.
In addition, the method of the proof will be somehow generalized to quantum mechanics 
in Section~\ref{sec:proof-prop-refprs}. 

Consider a classical Hamiltonian 
\begin{align}
H^{\mathrm{cl}}(x,y,\eta,\zeta)
=\tfrac12(\eta^2+\zeta^2)-x+q_1(x,y)
\ \ \text{for }
(x,y;\eta,\zeta)\in T^*\mathbb R^d\cong\mathbb R^{2d},
\label{eq:20062820}
\end{align}
where $q_1$ is from Condition~\ref{cond:one-body-starkPot}.
The associated Hamilton equations are 
\begin{align}
\dot x=\eta,\quad 
\dot y=\zeta,\quad
\dot\eta=1-\partial_xq_1,\quad
\dot\zeta=-\partial_yq_1.
\label{eq:200702}
\end{align}
First let us set $q_1\equiv 0$,
so that we can explicitly solve \eqref{eq:200702}.
A solution to \eqref{eq:200702} with initial data $(x_0,y_0;\eta_0,\zeta_0)\in T^*\mathbb R^d$ is given by 
\begin{align}\label{eq:freeClas}
x=\tfrac12t^2+t\eta_0+x_0,\quad
y=t\zeta_0+y_0,\quad
\eta=t+\eta_0,\quad
\zeta=\zeta_0.
\end{align}
Hence for this free Stark orbit we obviously have 
\begin{align}
 \label{eq:clasLarge0}
f-t=\mathcal O(1)\ \ \text{as }t\to\infty. 
\end{align}
This implies that 
we can think of the quantity
$f$ as an `effective time',
which allows us to effectively 
rewrite time-dependent observables by time-independent ones.
For example, the following bounds hold as $t\to\infty$
\begin{align}
 \label{eq:clasLarge}
\eta-f=\mathcal O(f^{-1}),\quad 
\zeta-g=\mathcal O(f^{-1}),\quad
f^{-1}g=\mathcal O(f^{-1}) .
 \end{align}
This says that $(f,g)$ is comparable to the momentum $(\eta,\zeta)$. 

Let us rephrase \eqref{eq:clasLarge} in terms of the classical `gamma observables'.
For simplicity consider only a zero-energy scattering orbit,
and set
\begin{align}\label{eq:gammaClas}
 \begin{split}
 \gamma^{\mathrm{ex}}&
=(\eta,\zeta)-\nabla \theta^{\mathrm{ex}},
\quad 
\tilde\gamma^{\mathrm{ex}}=f^{-1}g,
\quad 
\gamma^{\mathrm{ex}}_{\|}=(\nabla\theta^{\mathrm{ex}})\cdot \gamma^{\mathrm{ex}} ,
 \end{split}
\end{align} 
where $\theta^{\mathrm{ex}}$ is from \eqref{eq:20070210}. 
These are exact (i.e. unsimplified) classical versions of the quantum gamma observables 
\eqref{eq:QM1} and \eqref{eq:200820} with $\lambda=0$. 
We also note that $\Gamma^{\mathrm{ex}}=(\gamma^{\mathrm{ex}},\tilde\gamma^{\mathrm{ex}})$ is a $(2d-1)$-dimensional variable. 
Whence we might
think of $\Gamma^{\mathrm{ex}}$ as a complete set of variables on the zero-energy
shelf, here being
unjustified though.
Now we easily see from \eqref{eq:clasLarge} that 
\begin{align}
\gamma^{\mathrm{ex}}=\mathcal O(f^{-1}),\quad 
\tilde\gamma^{\mathrm{ex}}=\mathcal O(f^{-1}),
\label{eq:200907}
\end{align}
and further that by Lemma~\ref{lem:20062814}
\begin{align}
\label{eq:quad1b}
 \gamma_{\|}^{\mathrm{ex}}= H^{\mathrm{cl}} -\tfrac12(\gamma^{\mathrm{ex}})^2
=-\tfrac12(\gamma^{\mathrm{ex}})^2
=\mathcal O(f^{-2}).
\end{align} 
We remark, most importantly, that 
the last bound \eqref{eq:quad1b} is sharper than what 
 a priori can be read off from 
\eqref{eq:clasLarge} or \eqref{eq:200907}. A similar observation was
done in the context of Schr\"odinger operators with a long-range
potential in \cite{HS}.
This simple application of \eqref{eq:quad1b} provides a heuristics to Proposition~\ref{prop:phase-space-radiation}
why the part of $\gamma(\lambda)$ in the parallel direction accepts a doubled weight,
see also \eqref{eq:p3}.

The above properties should hold also for a nonzero perturbation $q_1\not\equiv 0$,
and next we are going to verify it. 
To be comparable with Proposition~\ref{prop:phase-space-radiation} and
the quantum arguments in Section~\ref{sec:proof-prop-refprs}
already at this stage
we introduce simplified classical gamma observables. 
Let $\theta_\lambda$ be given by \eqref{eq:Jac1} and set 
\begin{align*} 
\gamma=\gamma^{\mathrm{cl}}=(\eta,\zeta)-\nabla\theta_0,
\quad 
\tilde\gamma=\tilde\gamma^{\mathrm{cl}}=f^{-1}g,
\quad
\gamma_\|=\gamma^{\mathrm{cl}}_\|=2r(\nabla f)\cdot\gamma^{\mathrm{cl}}.
\end{align*} 
We drop at this point for convenience the superscript ${}^{\mathrm{cl}}$.

\begin{lemma}\label{lem:200909}
Let $(x,y,\eta,\zeta)$ be a zero-energy forward scattering orbit 
for the perturbed Stark Hamiltonian \eqref{eq:20062820} (by definition
a forward scattering orbit obeys
$x^2+y^2\to \infty\ \text{as }t\to+\infty$),
and define $\gamma=\gamma^{\mathrm{cl}},\tilde\gamma=\tilde\gamma^{\mathrm{cl}}$ and 
$\gamma_\|=\gamma_\|^{\mathrm{cl}}$ as above.
Then
for any $s\in (0,1+\delta)$, as $t\to +\infty$, 
\begin{align}
 \label{eq:Gampar} 
\gamma^2=\mathcal O\bigl(f^{-s}\bigr),\quad 
\tilde\gamma^2=\vO\parb{f^{-(s+2)/2}},\quad 
\gamma_{\|}=\vO\parb{f^{-s}}.
\end{align}
\end{lemma}
\begin{remark*}
We are going to use a minimal structure without explicit time parameter,
so that the method might appear easier to generalize to quantum
mechanics. In fact the proof of the quantum analogue
 Proposition \ref{prop:phase-space-radiation}, although being technically more complicated, will indeed follow the
same basic scheme.
\end{remark*}

\begin{proof}
First we note that, 
since $(x,y,\eta,\zeta)$ is a scattering orbit and $x$ is bounded below by conservation of energy, 
we can find $c_1>0$ such that for large $t\ge 0$
$${f\ge c_1|g|},\quad r+x> 2$$
along the orbit. 
This in turn implies that there exists $C_1>0$ such that 
for large $t\ge 0$
$$\tfrac12f^2\le r\le C_1f^2.$$
This is a classically allowed region, and we will use this bound
without mentioning.

Now by the formula
 \begin{align}
 \label{eq:ei}
 \tfrac12\abs{\nabla \theta_0}^2-x=\tfrac14 r^{-1}\abs{g}^4,
 \end{align} 
we can deduce the important identity 
\begin{align}
\label{eq:quad1}
0=H^{\mathrm{cl}}
=\tfrac12\gamma^2+\tfrac12f^2r^{-1}\gamma_\|+\tfrac14f^4r^{-1}\tilde\gamma^4+q_1.
\end{align}
cf.\ \eqref{eq:quad1b}.
First, by \eqref{eq:quad1} it suffices to show 
\begin{align}
\gamma_\|=\mathcal O(f^{-s}).
\label{eq:20090718}
\end{align}
To prove \eqref{eq:20090718},
letting $D=\tfrac{\mathrm d}{\mathrm dt}$ be the time-derivative, we compute
\begin{align}
D\bigl((f^s\gamma_\|)^2\bigr)
&=2sf^{2s-1}\gamma_\|^2(D f)
+2f^{2s}\gamma_\|\bigl(D\gamma_\|\bigr)
,
\label{eq:20090719}
\end{align}
and show that this quantity is almost negative. 
We bound the first term on the right-hand side of \eqref{eq:20090719}
as, by using \eqref{eq:quad1} twice and the Cauchy-Schwarz inequality, 
\begin{align*}
2sf^{2s-1}\gamma_\|^2(D f)
&=
2sf^{2s-1}\gamma_\|^2(\nabla f)\cdot(\eta,\zeta)
\\&
=
sf^{2s-1}r^{-1}\gamma_\|^3
+sf^{2s+1}r^{-1}\gamma_\|^2
\\&
\le 
sf^{2s+1}r^{-1}\gamma_\|^2
+
C_2f^{2s-4-\delta}\gamma_\|^2
\\&
\le 
sf^{2s+1}r^{-1}\gamma_\|^2
+C_3f^{2s-2-\delta}\gamma^2
\\&
\le 
sf^{s+1}r^{-1}\gamma_\|^2
-\tfrac12C_3f^{2s-\delta}r^{-1}\gamma_\|
+C_4f^{2s-3-2\delta}
\\&
\le 
(s+\epsilon)f^{2s+1}r^{-1}\gamma_\|^2
+C_5f^{2s-3-2\delta}.
\end{align*}
Here $\epsilon\in(0,1-s/2)$ is a fixed constant. 
As for the second term of \eqref{eq:20090719}, 
{we compute the last factor by using e.g.\ \eqref {eq:ort} and \eqref{eq:ei}} as 
\begin{align*}
\begin{split}
D\gamma_\|
&= 
2(\eta,\zeta)\cdot(\nabla r\nabla f)\gamma
+2r(\nabla f)\cdot \bigl((\nabla x)-(\nabla q_1)-(\nabla^2\theta_0)(\eta,\zeta)\bigr)
\\&
=
2\gamma\cdot(\nabla r\nabla f)\gamma
+\tfrac14fg^4r^{-2}
+\tfrac12f^3r^{-2}\gamma_\|
-fr^{-1}\gamma_\|
-2r(\nabla f)\cdot(\nabla q_1),
\end{split}
\end{align*} 
which implies by Lemma~\ref{lem:200821} that 
\begin{align*}
-fr^{-1}\gamma_\|
-C_6f^{-2-\delta}
\le D\gamma_\|
\le 
C_6 f^{-1}\bigl(
\gamma^2
+f^2\tilde\gamma^4
+|\gamma_\||
+f^{-1-\delta}
\bigr).
\end{align*}
Then the second term of \eqref{eq:20090719} is bounded as, 
by using \eqref{eq:quad1} and the Cauchy-Schwarz inequality again, 
\begin{align*}
2f^{2s}\gamma_\|\bigl(D\gamma_\|\bigr)
&
=
-f^{2s-2}r
\bigl(2\gamma^2+f^4r^{-1}\tilde\gamma^4\big)\bigl(D\gamma_\|\bigr)
-2q_1f^{2s-2}r
\bigl(D\gamma_\|\bigr)
\\&
\le 
f^{2s-1}\bigl(2\gamma^2+f^4r^{-1}\tilde\gamma^4\big)\gamma_\|
+C_7f^{2s-2-\delta}\bigl(\gamma^2+f^2\tilde\gamma^4\big)
\\&\phantom{{}={}}{}
+C_7f^{2s-2-\delta}\bigl(\gamma^2+f^2\tilde\gamma^4
+|\gamma_\||+f^{-1-\delta}\bigr)
\\&
\le 
-2f^{2s+1}r^{-1}\gamma_\|^2
+C_8f^{2s-2-\delta}\bigl(\gamma^2+f^2\tilde\gamma^4
+|\gamma_\||+f^{-1-\delta}\bigr)
\\&
\le 
-2f^{2s+1}r^{-1}\gamma_\|^2
+C_9f^{2s-2-\delta}|\gamma_\||
+C_9f^{2s-3-2\delta}
\\&
\le 
-(2-\epsilon)f^{2s+1}r^{-1}\gamma_\|^2
+C_{10}f^{2s-3-2\delta}.
\end{align*}
Now by the above computations we have \eqref{eq:20090719} bounded as 
\begin{align}
\begin{split}
D\bigl((f^s\gamma_\|)^2\bigr)
&\le 
-(2-s-2\epsilon) fr^{-1}(f^s\gamma_\|)^2
+C_{11}f^{2s-3-2\delta}
\\&
\le 
-c_2 f^{-1}
\bigl((f^s\gamma_\|)^2-C_{12}f^{2s-2-2\delta}\bigr)
,
\end{split}
\label{eq:20090913}
\end{align}
from which we can deduce the quantity $(f^s\gamma_\|)^2$ is bounded as $t\to+\infty$.
In fact, if not, we can find a sequence $(t_n)_{n\in\mathbb N}$ such that 
$$\lim_{n\to+\infty}t_n\to+\infty,\quad
(f^s\gamma_\|)^2(t_n)\ge n,\quad 
D\bigl((f^s\gamma_\|)^2\bigr)(t_n)\ge 0,
$$
but this contradicts \eqref{eq:20090913} since $f(t_n)\to+\infty$ as $n\to+\infty$ due to 
that the orbit is scattering. 
Therefore it follows that $(f^s\gamma_\|)^2$ is bounded as $t\to+\infty$, 
verifying \eqref{eq:20090718}. We are done.
\end{proof}
\begin{remark*}
From \eqref{eq:gammaClas} and the decomposition \eqref{eq:quad1} 
the natural simplified gamma observable in the parallel direction would be 
$$\gamma_\|'=(\nabla \theta_0)\cdot \gamma=f^2(\nabla f)\cdot \gamma
=\tfrac12f^2r^{-1}\gamma_\|.$$
However, the Hessian $\nabla f^2\nabla f=\nabla^2\theta_0$ made from its coefficients 
has only a weaker positivity than $2\mathop{\mathrm{Re}}(\nabla r\nabla f)$ 
made from those of $\gamma_\|$.
If we adopted $\gamma_\|'$ instead of $\gamma_\|$, 
we would need more on the classical orbits to prove the assertion. 
We also remark that we can not find a function having $2r\nabla f$ as its gradient,
since $\curl (2r\nabla f)\neq 0$.
\end{remark*}

\section{Reduction to radiation condition bounds}\label{sec:reduct-radi-cond-1}

In this section 
we prove our main results stated in Subsection~\ref{subsec:200825}, 
assuming the radiation condition bounds of Proposition~\ref{prop:phase-space-radiation}. 
The proof of Proposition~\ref{prop:phase-space-radiation} 
is postponed to Section~\ref{sec:proof-prop-refprs}. 
{We shall  }often use the notation $\chi(\cdot <\kappa)$ and $\chi^\perp(\cdot<\kappa)$
from \eqref{eq:20100311} and \eqref{eq:20100312}, respectively.

First we show that the limits of \eqref{eq:f_limit} exist for any $\psi\in C_{\mathrm c}^\infty(\R^d)$. 
By noting $\partial_f\ln(J^{-1/2})=\tfrac12\Div\,2r(\nabla f)$ we have 
\begin{align*}
p_f\vR^{\pm}_{\lambda}\psi 
&= 
-{\mathrm i}\parb{\partial_f\ln(J^{-1/2})}\vR^{\pm}_{\lambda}\psi
+\omega_\pm J^{-1/2}\mathrm e^{\mp {\mathrm i}\theta_\lambda}
\parb{2r(\nabla f)\cdot\gamma(\lambda)} R(\lambda\pm {\mathrm i}0)\psi
\\&= 
\omega_\pm J^{-1/2}\mathrm e^{\mp {\mathrm i}\theta_\lambda}\gamma_\|(\lambda)
R(\lambda\pm {\mathrm i}0)\psi, 
\end{align*}
see \eqref{eq:QM1} and \eqref{eq:200820} for $\gamma(\lambda)$ and $\gamma_\|(\lambda)$, respectively. 
Then by using the Cauchy-Schwarz inequality 
and Proposition~\ref{prop:phase-space-radiation}, 
we can estimate for $\epsilon\in(0, \delta)$ as 
\begin{align*}
 \begin{split}
 \int_2^\infty\|p_f\vR^\pm_\lambda\psi(f, \cdot)\|_\Sigma\,\d f 
&\le 
C_1\int_2^\infty \|f^{1/2+\epsilon}p_f\vR^\pm_\lambda\psi(f, \cdot)\|_\Sigma^2\,\d f 
\\&\le 
C_1\|f^{1/2+\epsilon}\gamma_\|(\lambda) R(\lambda\pm \i0)\psi\|_{L^2}^2 <\infty. 
 \end{split}
\end{align*}
The existence of \eqref{eq:f_limit} for $\psi\in C_{\mathrm c}^\infty(\R^d)$ follows from this estimate.

\begin{lemma}\label{lem:inner-prod-waveop}
Let $\lambda\in\R$. 
Then for any $\psi\in C_{\mathrm c}^\infty(\R^d)$ and $\xi\in C_{\mathrm c}^\infty(\R^{d-1})$ 
\begin{equation}\label{eq:2009212324}
\langle \vF^\pm(\lambda)\psi, \xi\rangle_\Sigma 
= 
\langle \psi, \phi^\pm_\lambda[\xi]\rangle 
- \langle R(\lambda\pm \i0)\psi, \psi^\pm_\lambda[\xi]\rangle. 
\end{equation}
In particular for any $\xi\in C_{\mathrm c}^\infty(\R^{d-1})$
\begin{equation}\label{eq:2009281723}
\vF^\pm(\lambda)^*\xi 
= \phi_\lambda^\pm[\xi]-R(\lambda\mp\i0)\psi_\lambda^\pm[\xi]\ (\in\vB^*).
\end{equation}
\end{lemma}
\begin{proof}
We consider only for the upper sign for notational simplicity. 
We let $\phi=R(\lambda+\i0)\psi$ and 
introduce $\chi_m=\chi_m(f)=\chi(f/2^m<1)$ for $m\in\N$. 
By using \eqref{eq:2009212345} we compute for any $m\in\N$
\begin{equation}
\begin{split}\label{eq:2009221446}
2\langle \psi, \chi_m\phi_\lambda^+[\xi]\rangle
&= 
\langle \bigl(B-(\partial_f\theta_\lambda)\bigr)(2r)^{-1}\bigl(B+(\partial_f\theta_\lambda)\bigr)\phi, \chi_m\phi_\lambda^+[\xi]\rangle 
\\&\phantom{{}={}}{} 
+\langle (L+2\vO(f^{-1-\delta}))\phi, \chi_m\phi_\lambda^+[\xi]\rangle 
\\&= 
-{\mathrm i}\langle (2r)^{-1}\bigl(B+(\partial_f\theta_\lambda)\bigr)\phi, \chi_m'\phi_\lambda^+[\xi]\rangle 
\\&\phantom{{}={}}{} 
+\langle (2r)^{-1}\bigl(B+(\partial_f\theta_\lambda)\bigr)\phi, \chi_m\bigl(B-(\partial_f\theta_\lambda)\bigr)\phi_\lambda^+[\xi]\rangle 
\\&\phantom{{}={}}{} 
+\langle \phi, \chi_m(L+2\vO(f^{-1-\delta}))\phi_\lambda^+[\xi]\rangle 
\\&= 
-\i\int_1^\infty \chi_m'\langle
(2r)^{-1}2(\partial_f\theta_\lambda)\phi,
J^{-1}\phi_\lambda^+[\xi]\rangle_\Sigma\,\d f
\\&\phantom{{}={}}{} 
-\i\int_1^\infty \chi_m'\langle (2r)^{-1}\bigl(B-\partial_f\theta_\lambda\bigr)\phi, J^{-1}\phi_\lambda^+[\xi]\rangle_\Sigma\,\d f
\\&\phantom{{}={}}{} 
+\langle (2r)^{-1}\bigl(B+\partial_f\theta_\lambda\bigr)\phi, \chi_m\bigl(B-\partial_f\theta_\lambda\bigr)\phi_\lambda^+[\xi]\rangle 
\\&\phantom{{}={}}{} 
+{\langle \phi, \chi_m\parb{L+2\vO(f^{-1-\delta})}\phi_\lambda^+[\xi]\rangle.} 
\end{split}
\end{equation} 

The first term to the right converges to $2\langle
\vF^+(\lambda)\psi, \xi \rangle_\Sigma $ for $m\to\infty$. The second
term 
vanishes in the limit due to
Proposition~\ref{prop:phase-space-radiation}. For the third term we
can safely take the factor
$(2r)^{-1}\bigl(B+\partial_f\theta_\lambda\bigr)$ to the other side,
with vanishing error (from commuting with $\chi_m$) in the limit due to \eqref{eq:200816}. 
Thus we conclude by taking the limit on both sides of
\eqref{eq:2009221446} that 
\begin{align*}
2\langle \psi, \phi_\lambda^+[\xi]\rangle
&= 
2\langle \vF^+(\lambda)\psi, \xi \rangle_\Sigma 
+\langle \phi, (B+{\partial_f\theta_\lambda})(2r)^{-1}(B-\partial_f\theta_\lambda)\phi_\lambda^+[\xi]\rangle 
\\&\phantom{{}={}}{} 
+{\langle \phi, \parb{L+2\vO(f^{-1-\delta})}\phi_\lambda^+[\xi]\rangle.} 
\end{align*} 
Whence
\begin{equation*}
\langle \vF^+(\lambda)\psi, \xi \rangle_\Sigma 
= 
\langle \psi, \phi_\lambda^+[\xi]\rangle
- \langle R(\lambda+{\mathrm i}0)\psi, \psi_\lambda^+[\xi]\rangle, 
\end{equation*}
and we are done.
\end{proof}

\begin{proof}[Proof of Theorem~\ref{thm:dist-four-transf}]
We prove (1), (2) and (4). 
The assertion (3) will be proved at the end of the proof of Theorem~\ref{thm:char-gener-eigenf-1}. 

Note that the right-hand side of \eqref{eq:extbF2}
 exists for $\psi\in C_{\mathrm c}^\infty(\R^d)$ since this assertion
 is weaker than the existence of \eqref{eq:f_limit}. 
 For any $\psi\in\vB$ we note the bound
\begin{equation*}
\sup_{\rho>2}
\left\|{-\!\!\!\!\!\!\int_\rho} \vR^\pm_\lambda\psi(f, \cdot)\,\d f \right\|_\Sigma 
\le 
C \|\psi\|_\vB, 
\end{equation*}
with the constant $C>0$ being locally uniform in $\lambda\in\R$ 
and independent of $\psi\in\vB$. 
Then by a density argument we conclude that the limits in \eqref{eq:extbF2}
 exist for any $\psi\in\vB$. This is (1). 

Let us consider (4) before verifying (2). 
We note we may assume $\psi\in C_{\mathrm c}^\infty(\R^d)$. 
By integration by parts and using Proposition~\ref{prop:phase-space-radiation} 
we compute with $\phi=R(\lambda\pm{\mathrm i}0)\psi$ and $\chi_m=\chi(f/2^m<1)$ 
\begin{align*}
2\pi\langle \psi, \delta(H-\lambda)\psi\rangle 
&= 
\pm2\Im\langle(H-\lambda)\phi, \phi\rangle 
\\&= 
\mp\lim_{m\to\infty}\Re \langle (\nabla f\cdot p)\phi, \chi_m'\phi\rangle
\\&= 
\mp\lim_{m\to\infty}\Re \langle \nabla f\cdot(\gamma(\lambda)\pm \nabla\theta_\lambda)\phi, \chi_m'\phi\rangle
\\&= 
\lim_{m\to\infty}\Re
 \langle \parb{\mp(2r)^{-1}\gamma_\|(\lambda)+\tfrac12f^2r^{-1}}\phi,
 \chi_m'\phi\rangle
\\&= 
\lim_{m\to\infty} \langle \parb{1- \tfrac{g^2}{f^2+g^2}}\phi, \chi_m'\phi\rangle
\\&= 
2\pi\|\vF^\pm(\lambda)\psi\|_\Sigma^2-\lim_{m\to\infty} \langle
 \tfrac{{\tilde\gamma}^2}{1+{\tilde\gamma}^2}\phi,
 \chi_m'\phi\rangle
\\&= 
2\pi\|\vF^\pm(\lambda)\psi\|_\Sigma^2.
\end{align*}
Thus we have proved (4). 

Joint continuity of $\vF^\pm(\lambda)\psi\in \Sigma$ in
 $\lambda\in\R$ and $\psi\in\vB$ 
 reduces by \eqref{eq:fund} to the assertion that for any fixed
 $\psi\in C_{\mathrm c}^\infty(\R^d)$ the maps $\lambda\to
 \vF^\pm(\lambda)\psi$ are continuous.
 Next by combining \eqref{eq:fund} with the strong weak-star
 continuity of $\delta(H-\lambda)$ it remains to show that $\lambda\to
 \langle \vF^\pm(\lambda)\psi, \xi\rangle_\Sigma $ are continuous for
 any fixed $\xi\in C_{\mathrm c}^\infty(\R^{d-1})$. These (weak)
 continuity assertions follow readily from the representations \eqref{eq:2009212324}. 
Hence (2) is proved.
\end{proof}

Now let us prove the following lemma 
which is a part of Proposition~\ref{cor:scatt-matr-gener}. 
\begin{lemma}\label{lem:2010272050}
For any $\xi \in C^\infty_{\mathrm c}(\R^{d-1})$
 \begin{subequations}
\begin{align}\label{eq:rep1}
 \begin{split}
 \vF^\pm(\lambda)^* \xi
= \phi_\lambda^\pm[\xi]-R(\lambda\mp\i 0)\psi_\lambda^\pm[\xi]
=2\pi\i \delta(H-\lambda)\psi_\lambda^\pm[\xi],
\end{split}
 \end{align}
and (in particular) for any such $\xi$ 
\begin{align}\label{eq:rep3}
 \xi= \pm 2\pi\i \vF^\pm(\lambda)\psi_\lambda^\pm[\xi]. 
 \end{align}
\end{subequations} 
\end{lemma}
\begin{proof}
The first equality of \eqref{eq:rep1} {is equivalent to
 Lemma~\ref{lem:inner-prod-waveop}. 
By the Sommerfeld's uniqueness result \cite[Corollary~2.15]{AIIS1} it
follows that 
\begin{equation}\label{eq:2010271869}
\phi_\lambda^\pm[\xi]=R(\lambda\pm\i0)\psi_\lambda^\pm[\xi].
\end{equation} We substitute \eqref{eq:2010271869} in the middle expression, yielding
the second 
 equality of \eqref{eq:rep1}.
We deduce from \eqref{eq:2010271869} that 
\begin{equation*}
\chi^\perp(f<2)\xi
=\omega_\pm J^{-1/2}{\mathrm e}^{\mp\i\theta_\lambda}R(\lambda\pm\i0)\psi_\lambda^\pm[\xi]
=\vR^\pm_\lambda\psi_\lambda^\pm[\xi].
\end{equation*}
Thus by taking an average limit $\vSigmalim_{\rho\to \infty}-\!\!\!\!\!\int_\rho \cdot\,\d f$ 
on the both side we conclude by \eqref{eq:extbF2} that}
\begin{equation*}
\xi
=\vSigmalim_{\rho\to \infty}-\!\!\!\!\!\!\int_\rho \chi^\perp(f<2)\xi\,\d f
=\vSigmalim_{\rho\to \infty}-\!\!\!\!\!\!\int_\rho \vR^\pm_\lambda\psi_\lambda^\pm[\xi]\,\d f
=\pm2\pi \i \vF^\pm(\lambda)\psi_\lambda^\pm[\xi].
\end{equation*}
Thus we have \eqref{eq:rep3}. 
\end{proof} 

{By \eqref{eq:rep3}} 
the ranges $\set{\vF^\pm(\lambda)\psi|\, \psi\in \vB}$ are dense in
$\Sigma$ for any $\lambda$. This fact in combination with
Theorem~\ref{thm:dist-four-transf}~\eqref{item:delta} allows us to
{define} $S(\lambda)$ as the unique unitary operator on $\Sigma$ obeying 
\begin{align}\label{eq:S}
\vF^+(\lambda)\psi =S(\lambda)\vF^-(\lambda)\psi\ \ \text{for all }\
\psi\in\vB.
\end{align}
Now we can complete the proof of Proposition ~\ref{cor:scatt-matr-gener}. 
\begin{proof}[Proof of Proposition~\ref{cor:scatt-matr-gener}]
Let $\xi\in C_{\mathrm c}^\infty(\R^{d-1})$. 
{
The formulas \eqref{eq:rep1-} and \eqref{eq:rep3-} are
treated above. 
 Using 
\eqref{eq:rep3} and \eqref{eq:S} we compute }
\begin{equation*}
S(\lambda)\xi
=-2\pi\i S(\lambda)\vF^-(\lambda)\psi_\lambda^-[\xi]
=-2\pi\i\vF^+(\lambda)\psi_\lambda^-[\xi]
=-\vSigmalim_{\rho\to\infty} \intR \vR_\lambda^+\psi_\lambda^-[\xi](f,\cdot)\,\d f.
\end{equation*}
This is \eqref{eq:formS1}. 
By substituting $S(\lambda)\xi=-2\pi\i\vF^+(\lambda)\psi_\lambda^-[\xi]$ into the left-hand side of \eqref{eq:formS2} 
and using \eqref{eq:2009281723} we obtain 
\begin{align*}
\tfrac1{2\pi \i}\inp{{\xi'},S(\lambda)\xi}
&= 
\inp{\xi', -\vF^+(\lambda)\psi_\lambda^-[\xi]}
\\&= 
\inp{-\vF^+(\lambda)^*\xi', \psi_\lambda^-[\xi]}
\\&= 
\inp{R(\lambda-\i0)\psi_\lambda^+[\xi']-\phi_\lambda^+[\xi'], \psi_\lambda^-[\xi]}
\\&= 
\inp{\psi_\lambda^+[{\xi'}], R(\lambda+\i0)\psi_\lambda^-[\xi]}
-\inp{\phi_\lambda^+[{\xi'}],\psi_\lambda^-[\xi]}.
\end{align*}
This is \eqref{eq:formS2}.
\end{proof}

\begin{proof}[Proof of Theorem~\ref{thm:unit-equiv}]
We consider only the upper sign case. The proof
 essentially mimic similar
arguments in \cite{ACH}.

\vspace{1mm}
\noindent
\emph{Step I.} 
First we show that $\vF^+$ can be extended to an isometry $\vH\to\widetilde\vH$ 
 satisfying $\vF^+H\subseteq M_\lambda\vF^+$. 
By Stone's formula and \eqref{eq:fund} we have 
\begin{equation*}
\|\vF^+\psi\|_{\widetilde\vH}=\|\psi\|_\vH\ \ \text{for any }\ \psi\in\vB.
\end{equation*}
Thus the former assertion is verified. 
For the latter assertion, it suffices to show that for any $\psi\in\vB$
\begin{equation*}
\vF^+(H-{\mathrm i})^{-1}\psi=(M_\lambda-{\mathrm i})^{-1}\vF^+\psi.
\end{equation*}
Noting the resolvent equation 
\begin{equation*}
R(\lambda+{\mathrm i}0)R({\mathrm i})
=(\lambda-{\mathrm i})^{-1}R(\lambda+{\mathrm i}0)
-(\lambda-{\mathrm i})^{-1}R({\mathrm i}), 
\end{equation*}
we can compute 
\begin{align*}
\vF^+(\lambda)R({\mathrm i})\psi 
&= 
(2\pi{\mathrm i})^{-1}\,\vSigmalim_{\rho\to\infty}{-\!\!\!\!\!\!\int_\rho} 
(\lambda-{\mathrm i})^{-1}\vR^\pm_\lambda\psi(f,\cdot)\,\d f 
\\&\phantom{{}={}}{}
-(2\pi{\mathrm i})^{-1}\,\vSigmalim_{\rho\to\infty}{-\!\!\!\!\!\!\int_\rho} 
(\lambda-{\mathrm i})^{-1}\omega_+J^{-1/2}\mathrm e^{-{\mathrm i}\theta_\lambda}
R({\mathrm i})\psi(f, \cdot)\,\d f 
\\&= 
(\lambda-{\mathrm i})^{-1}\vF^+(\lambda)\psi. 
\end{align*}

\vspace{1mm}
\noindent
\emph{Step I\hspace{-1pt}I.} 
We prove $\vF^+: \vH\to\widetilde\vH$ is surjective. 
In combination with Step I it then follows that $\vF^+$ is unitary 
and that $\vF^+H=M_\lambda\vF^+$. 

By arguing as in Step I and invoking the
 Helffer-Sj\"ostrand formula, see \eqref{eq:hsF}, we obtain that for all $h\in C_{\mathrm c}^\infty(\R)$ and $\psi\in\vB$ 
\begin{equation*}
\vF^+(\lambda)h(H)\psi=h(\lambda)\vF^+(\lambda)\psi. 
\end{equation*} 
(The
fact that $h(H)\psi\in
\vB$ follows from \cite[Theorem 14.1.4]{Ho}.) 
Now suppose that $g(\cdot)\in\Ker\,(\vF^+)^*\subseteq\widetilde\vH$. 
Then we have for any $h\in C_{\mathrm c}^\infty(\R)$ and $\psi\in\vB$
\begin{equation*}
\int_\R h(\lambda)\langle g(\lambda), \vF^+(\lambda)\psi\rangle_\Sigma\,\d\lambda
= 
\langle (\vF^+)^*g(\cdot), h(H)\psi\rangle_\vH=0. 
\end{equation*}
This implies 
\begin{equation*}
\langle g(\lambda), \vF^+(\lambda)\psi\rangle_\Sigma=0\ \ \text{for a.e. }\lambda\in\R.
\end{equation*}
Let $\set{\psi_k}$ be any countable dense subset of
 $\vB$. Then by the density 
of $\{\vF^+(\lambda)\psi_k\}\subseteq\Sigma$
for any $\lambda$, 
we see that $g(\lambda)=0$ almost everywhere. So $g(\cdot)=0$. 
Hence we have $\Ker\,(\vF^+)^*=\{0\}$, and then $\Ran\,\vF^+=\widetilde\vH$. 
\end{proof}

We introduce several lemmas as a preparation for the proof of Theorem~\ref{thm:char-gener-eigenf-1}. 
First we verify a uniqueness assertion.
\begin{lemma}\label{lem:uniquely-determine}
Suppose that $\xi_\pm\in\Sigma$ and $\phi\in\vE_\lambda$ satisfy 
\begin{align}\label{eq:asphi}
\phi -\phi_\lambda^+[\xi_+]-\phi_\lambda^-[\xi_-]\in\vB_0^*.
\end{align}
Then one has 
\begin{subequations}
\begin{align}
\|\xi_+\|^2+\|\xi_-\|^2 
&= 2\pi \lim_{n\to\infty} {2^{-n}\int_{2^n\le f<2^{n+1}}\,|\phi|^2\,\d x\,\d y}, 
\label{eq:2009280224}
\\
\|\xi_+\| &= \|\xi_-\|. \label{2009280225}
\end{align} 
In particular any two of the quantities in $\{\xi_-, \xi_+,
\phi\}$ uniquely determines the third one. 
\end{subequations}
\end{lemma}

\begin{proof}
We can compute the right-hand side of \eqref{eq:2009280224} as
\begin{align*}
&\quad\hspace{-3mm}
2\pi \lim_{n\to\infty} 2^{-n}\int_{2^n\le f<2^{n+1}}\,|\phi|^2\,\d x\,\d y 
\\&= 
2\pi \lim_{n\to\infty} 2^{-n}\int_{2^n\le f<2^{n+1}}\,
\left|\phi_\lambda^+[\xi_+]+\phi_\lambda^-[\xi_-]\right|^2\,\d x\,\d y
\\&= {
\|\xi_+\|_\Sigma^2+\|\xi_-\|_\Sigma^2
+2\Re\,\lim_{n\to\infty} 2^{-n}\int_{2^n\le f<2^{n+1}}\,\mathrm e^{{\mathrm i}\pi d/2}
\mathrm e^{-2{\mathrm i}\theta_\lambda}
\langle \xi_+, \xi_-\rangle_\Sigma\,\d f.}
\end{align*} 
For the last term we substitute $\mathrm e^{-2{\i}\theta_\lambda}
=-(2{\mathrm i}(f^2+\lambda))^{-1}\mathrm \partial_f{\mathrm e}^{-2{\i}\theta_\lambda}$, 
integrate by parts and then conclude that the limit vanishes. 
Thus we have \eqref{eq:2009280224}. 

Next we prove \eqref{2009280225}. 
We let $A=\Re\bigl((\nabla f)\cdot p\bigr)$ and $\chi_n=\chi(f/2^n<1)$ for $n\in\N$. 
Then by noting $AR({\mathrm i})\in\vL(\vB^*)$ 
and $A\phi=(\lambda-{\mathrm i})AR({\mathrm i})\phi\in\vB^*$ we have
\begin{align*}
0 = 
\lim_{n\to\infty}\langle\phi, {\mathrm i}[H, \chi_n]\phi\rangle 
&= 
\lim_{n\to\infty}\langle\phi, A\chi_n'\phi\rangle 
\\&= 
\lim_{n\to\infty}\langle A\phi, 
\chi_n'\bigl(\phi_\lambda^+[\xi_+]+\phi_\lambda^-[\xi_-]\bigr)\rangle 
\\&= 
\lim_{n\to\infty}\langle \phi, 
\chi_n'\bigl(A\phi_\lambda^+[\xi_+]+A\phi_\lambda^-[\xi_-]\bigr)\rangle 
\\&= 
\lim_{n\to\infty}\langle \phi, 
\tfrac{f^2+\lambda}{2r}\chi_n' 
\bigl(\phi_\lambda^+[\xi_+]-\phi_\lambda^-[\xi_-]\bigr)\rangle 
\\&= 
\tfrac1{2\pi}\left(\|\xi_-\|^2 - \|\xi_+\|^2\right).
\end{align*}
For the last step we first substitute
 \eqref{eq:asphi} to the left and note
that then we can replace the factor $ \tfrac{f^2+\lambda}{2r}$ by one
(thanks to Lebesgue's dominated convergence theorem). By the integration by
parts argument from above we then conclude \eqref{2009280225}.

Finally, the uniqueness statement follows from \eqref{eq:2009280224}, \eqref{2009280225}, 
linearity of $\phi_\lambda^\pm[\cdot]$ and the version of Rellich's theorem 
 \cite[ Theorem~2.5]{AIIS1}.
\end{proof}

Next we examine $\phi\in\vE_\lambda$ given by the formula
\eqref{eq:2009281723} with $\xi=\xi_-\in C_{\mathrm c}^\infty(\R^{d-1})$.
\begin{lemma}\label{lem:phi-from-xi-minus}
For any $\xi_-\in C_{\mathrm c}^\infty(\R^{d-1})$ we define $\phi\in\vE_\lambda$ 
and $\xi_+\in\Sigma$ by 
\begin{equation*}
\phi=\vF^-(\lambda)^*\xi_-=\phi_\lambda^-[\xi_-]-R(\lambda+{\mathrm i}0)\psi_\lambda^-[\xi_-],\quad 
\xi_+=-2\pi{\mathrm i}\vF^+(\lambda)\psi_\lambda^-[\xi_-].
\end{equation*}
Then \eqref{eq:gen1} and \eqref{eq:aEigenfw} hold for $\{\xi_-, \xi_+, \phi\}$.
\end{lemma}

\begin{proof}
We compute 
\begin{align*}
\phi -\phi_\lambda^+[\xi_+]-\phi_\lambda^-[\xi_-] 
&= 
2\pi{\mathrm i}\,\phi_\lambda^+\bigl[\vF^+(\lambda)\psi_\lambda^-[\xi_-]\bigr] 
- R(\lambda+{\mathrm i}0)\psi_\lambda^-[\xi_-] 
\\&= 
2\pi{\mathrm i}\,\omega_+^{-1}J^{1/2}\mathrm e^{{\mathrm i}\theta_\lambda}
\bigl( \vF^+(\lambda)\psi_\lambda^-[\xi_-]
- (2\pi{\mathrm i})^{-1}\vR_\lambda^+\psi_\lambda^-[\xi_-] \bigr). 
\end{align*}
Then by noting \eqref{eq:f_limit}
for $\psi\in C_{\mathrm c}^\infty(\R^d)$ 
 a simple approximation argument shows that the right-hand side belongs to $\vB_0^*$.
Thus we have \eqref{eq:gen1}. The assertions 
\eqref{eq:aEigenfw} follow from 
\eqref{eq:rep3}
and the definition of $S(\lambda)$.
\end{proof}

By similar arguments we obtain the following lemma (the proof is omitted).
\begin{lemma}\label{lem:phi-from-xi-plus}
For any $\xi_+\in C_{\mathrm c}^\infty(\R^{d-1})$ we define $\phi\in\vE_\lambda$ 
and $\xi_-\in\Sigma$ by 
\begin{equation*}
\phi=\vF^+(\lambda)^*\xi_+=\phi_\lambda^+[\xi_+]-R(\lambda-{\mathrm i}0)\psi_\lambda^+[\xi_+],\quad 
\xi_-=2\pi{\mathrm i}\vF^-(\lambda)\psi_\lambda^+[\xi_+].
\end{equation*}
Then \eqref{eq:gen1} and \eqref{eq:aEigenfw} hold for $\{\xi_-, \xi_+, \phi\}$.
\end{lemma}

Let us construct $\xi_\pm\in\Sigma$ from $\phi\in\vE_\lambda$.
\begin{lemma}\label{lem:xi-from-phi}
For any $\phi\in\vE_\lambda$ there exist $\xi_\pm\in\Sigma$ 
such that \eqref{eq:aEigenfw} hold.
\end{lemma}

\begin{proof}
By the definition of $S(\lambda)$ 
it suffices to show that there exists $\xi\in\Sigma$ 
such that $\phi=\vF^+(\lambda)^*\xi$ holds.

Let us take and fix a smooth cut-off function $\eta\in C_{\mathrm c}^\infty(\R)$ 
which satisfies $\eta(t)=t$ near $t=\lambda$. 
We introduce $\phi_\pm\in\vB^*$ and $\xi_n\in\Sigma$, $n\in\N$, by
\begin{equation*}
\phi_\pm=\dfrac1{2a_0}\chi^\perp_m(A\pm a_0)\phi,\quad 
\xi_n=2\pi{\mathrm i}\,\vF^+(\lambda)\chi_n(\eta(H)-\lambda)\phi_+, 
\end{equation*}
where {$A=\Re\bigl((\nabla f)\cdot p\bigr)$}, 
$a_0=\sqrt{(r+\lambda)/r}, \,\chi_m^\perp=1-\chi_m$ 
and $\chi_m$ is given as in the proof of
Lemma~\ref{lem:inner-prod-waveop} and used only with a fixed large
$m\in\N$.
Recalling from \cite{AIIS1} that $A$
is $H$-bounded indeed $\phi_\pm\in\vB^*$.
 We represent $\eta(H)$ by the Helffer-Sj\"ostrand formula 
\begin{align}\label{eq:hsF}
\eta(H) = \int_\C R(z)\,\d\mu(z);\quad 
\d\mu(z)=-(2\pi{\mathrm i})^{-1}\bar\partial_z\tilde\eta(z)\,\d z\,\d\bar z.
\end{align}

First we verify that the sequence $\{\xi_n\}\subseteq\Sigma$ is bounded. 
Noting $\vF^+(\lambda)(\eta(H)-\lambda)=0$ 
we represent 
 for any $v\in C_{\mathrm c}^\infty(\R^{d-1})$ with $\|v\|_\Sigma=1$, 
\begin{equation*}
\langle v,\xi_n\rangle_\Sigma 
= 
2\pi{\mathrm i}\langle \vF^+(\lambda)^*v, 
[\chi_n, \eta(H)]\phi_+\rangle_{\vB^*\times\vB}. 
\end{equation*}
Writing first 
\begin{align}\label{eq:2009292011}
 [\chi_n, \eta(H)]\phi_+ 
= 
-{\mathrm i}\int_\C R(z)(A\chi_n'+\tfrac{{\mathrm i}}2|\nabla f|^2\chi_n'')R(z)\,\d\mu(z)\phi_+,
\end{align} we can represent
\begin{align*}
 \langle \vF^+(\lambda)^*v, 
[\chi_n, \eta(H)]\phi_+\rangle_{\vB^*\times\vB}&=\langle
 \vF^+(\lambda)^*v, \parb{A\chi_n'+\tfrac{{\mathrm
 i}}2|\nabla
 f|^2\chi_n''}\eta_\lambda(H)
\phi_+\rangle_{\vB^*\times\vB};\\
&\eta_\lambda(t)=-{\mathrm i}\int_\C (\lambda-z)^{-1}(t-z)^{-1}\,\d\mu(z).
\end{align*} Noting that $\eta_\lambda(H)$ is a bounded operator on
$\vB^*$ we can then bound
\begin{align*}
 |\langle v,\xi_n\rangle_\Sigma |\leq
 C_1\parb{\|A\vF^+(\lambda)^*v\|_{\vB^*}+\|\vF^+(\lambda)^*v\|_{\vB^*}}\,\|\eta_\lambda(H)\phi_+\|_{\vB^*}\leq C_2\|v\|_\Sigma.
\end{align*} Since this is uniform in $n$, the sequence $\{\xi_n\}\subseteq\Sigma$ is bounded.

Now we choose a weakly convergent subsequence of $\{\xi_n\}$
and denote its weak limit by $\xi$. 
By changing notation we may assume 
$\mathop{\mathrm{w\text{-}lim}}_{n\to\infty}\xi_n=\xi\in\Sigma$. 
We claim that for this $\xi$ indeed $\vF^+(\lambda)\xi=\phi$. 
We compute with $\eta_\lambda(t):=(\eta(t)-\lambda)(t-\lambda)^{-1}$
\begin{align*}
&\quad\hspace{-3mm}\vF^+(\lambda)^*\xi
\\&= 
\mathop{\mathrm{w^*\text{-}\vB^*\text{-}lim}}_{n\to\infty} 
2\pi\i\vF^+(\lambda)^*\vF^+(\lambda)\chi_n(\eta(H)-\lambda)\phi_+
\\&= 
\mathop{\mathrm{w^*\text{-}\vB^*\text{-}lim}}_{n\to\infty} 
\left(R(\lambda+\i0)-R(\lambda-\i0)\right)\chi_n(\eta(H)-\lambda)\phi_+
\\&= 
\mathop{\mathrm{w^*\text{-}\vB^*\text{-}lim}}_{n\to\infty} 
\bigl(\eta_\lambda(H)\chi_n\phi 
+R(\lambda+\i0)[\chi_n, \eta(H)]\phi_+
+R(\lambda-\i0)[\chi_n, \eta(H)](\phi-\phi_+)\bigr)
\\&= 
\eta_\lambda(H)\phi 
+ \mathop{\mathrm{w^*\text{-}\vB^*\text{-}lim}}_{n\to\infty} 
\bigl(R(\lambda+\i0)[\chi_n, \eta(H)]\phi_+
+R(\lambda-\i0)[\chi_n, \eta(H)](\phi-\phi_+)\bigr).
\end{align*}
The first term is equal to $\phi$. 
By radiation condition bounds of \cite{AIIS1} {(or alternatively by the
stronger versions of 
Proposition \ref{prop:phase-space-radiation})} and \eqref{eq:2009292011}, 
we have 
\begin{align*}
&\quad\hspace{-3mm}
\mathop{\mathrm{w^*\text{-}\vB^*\text{-}lim}}_{n\to\infty}
R(\lambda+\i0)[\chi_n, \eta(H)]\phi_+
\\&= 
-\i \mathop{\mathrm{w^*\text{-}\vB^*\text{-}lim}}_{n\to\infty}
\int_\C R(z)R(\lambda+\i0)A\chi_n'R(z)\,\d\mu(z)\phi_+
\\&= 
\i \mathop{\mathrm{w^*\text{-}\vB^*\text{-}lim}}_{n\to\infty}
\int_\C R(z)R(\lambda+\i0)a_0\chi_n'R(z)\,\d\mu(z)\phi_+
\\&= 
-\tfrac{\i}2 \mathop{\mathrm{w^*\text{-}\vB^*\text{-}lim}}_{n\to\infty}
\eta'(H)R(\lambda+\i0)\chi_n'(A+a_0)\phi
\\&= 
-\tfrac{\i}2 \mathop{\mathrm{w^*\text{-}\vB^*\text{-}lim}}_{n\to\infty}
\eta'(H)R(\lambda+\i0)(A+a_0)\chi_n'\phi
=0.
\end{align*}
Similarly, by noting $\phi-\phi_+=\chi_m\phi-\phi_-$, 
\begin{align*}
&\quad\hspace{-3mm}
\mathop{\mathrm{w^*\text{-}\vB^*\text{-}lim}}_{n\to\infty} 
R(\lambda-\i0)[\chi_n, \eta(H)](\phi-\phi_+) 
\\&= 
-\tfrac{\i}2 \mathop{\mathrm{w^*\text{-}\vB^*\text{-}lim}}_{n\to\infty}
\eta'(H)R(\lambda-\i0)(A-a_0)\chi_n'\phi =0. 
\end{align*}
Hence the equality $\vF^+(\lambda)^*\xi=\phi$ is shown.
\end{proof}

\begin{proof}[Proof of Theorem~\ref{thm:char-gener-eigenf-1}]
For any $\xi_-\in\Sigma$ 
we choose a sequence $\{\xi_{-,n}\}\subseteq C_{\mathrm c}^\infty(\R^{d-1})$ 
satisfying $\xi_{-, n}\to\xi_-$ in $\Sigma$ as $n\to\infty$. 
By Lemma~\ref{lem:phi-from-xi-minus} we have 
\begin{equation*}
\vF^-(\lambda)^*\xi_{-,n}-\phi_\lambda^+[S(\lambda)\xi_{-,n}]-\phi_\lambda^-[\xi_{-,n}]\in\vB_0^*.
\end{equation*}
Then by the continuity of $\vF^-(\lambda)^*, S(\lambda)$ 
and $\phi_\lambda^\pm[\cdot]$ it follows, by taking the limit $n\to\infty$, that
\begin{equation*}
\vF^-(\lambda)^*\xi_--\phi_\lambda^+[S(\lambda)\xi_-]-\phi_\lambda^-[\xi_-]\in\vB_0^*.
\end{equation*}
Thus we have \eqref{eq:gen1} and \eqref{eq:aEigenfw} when $\xi_-$ is given first. 
By using Lemma~\ref{lem:phi-from-xi-plus} we can argue similarly when $\xi_+$ is given first. 
Therefore by Lemmas~\ref{lem:xi-from-phi} 
and~\ref{lem:uniquely-determine} we have shown (1).

To prove (2) it remains to show \eqref{eq:aEigenfwB}. Let $f_{-2}=\parbb{f\sqrt{f^2+g^2}}^{-1}$.
 By \eqref{eq:2009281723} 
we have for any $\xi\in C_{\mathrm c}^\infty(\R^{d-1})$
\begin{align}
\label{eq:2010081635}
\quad\hspace{-3mm}
\bigl(f_{-2}p_f\pm1\bigr)\vF^\pm(\lambda)^*\xi = 
\bigl(f_{-2}p_f\pm1\bigr)\phi_\lambda^\pm[\xi] 
-\bigl(f_{-2}p_f\pm1\bigr)R(\lambda\mp\i0)\psi_\lambda^\pm[\xi]
\end{align}
The first term on the right-hand side of \eqref{eq:2010081635} can be
written 
(by \eqref{eq:2008075b}) as
\begin{align*}
&\quad\hspace{-3mm}
\bigl(f_{-2}p_f\pm1\bigr)\phi_\lambda^\pm[\xi] 
\\&= 
f_{-2}
(B\mp\partial_f\theta_\lambda)\phi_\lambda^\pm[\xi]
\pm 2\phi_\lambda^\pm[\xi]
\\&\phantom{{}={}}{}
+f_{-2}
\Bigl(\i\mathop{\mathrm{div}}(r\nabla f)\pm\partial_f\theta_\lambda 
\mp f\sqrt{f^2+g^2}\Bigr)\phi_\lambda^\pm[\xi]. 
\end{align*}
By \eqref{eq:200816} and compactness of the support of $\xi$, 
we see that the first and the third terms belong to $\vB_0^*$. 
As for the second term on the right-hand side of \eqref{eq:2010081635}, we rewrite 
\begin{align*}
&\quad\hspace{-3mm}
-\bigl(f_{-2}p_f\pm 1\bigr)R(\lambda\mp\i0)\psi_\lambda^\pm[\xi].
\\&= 
-f_{-2}
(B \pm\partial_f\theta_\lambda)R(\lambda\mp\i0)\psi_\lambda^\pm[\xi]
- \i\mathop{\mathrm{div}}(r\nabla f)f_{-2}
R(\lambda\mp\i0)\psi_\lambda^\pm[\xi] 
\\&\phantom{{}={}}{}
\pm f_{-2}
\Bigl(\partial_f\theta_\lambda - f\sqrt{f^2+g^2}\Bigr)R(\lambda\mp\i0)\psi_\lambda^\pm[\xi].
\end{align*}
By Proposition~\ref{prop:phase-space-radiation} 
the first term on the right-hand side belongs to $\vB_0^*$. 
Moreover since $\mathop{\mathrm{div}}(r\nabla f)=\vO(f^{-1})$, also 
the second term belongs to $\vB_0^*$. 
Let us evaluate the third term. 
Introducing
$f_0=f_{-2}\bigl(\partial_f\theta_\lambda -
f\sqrt{f^2+g^2}\bigr)$, we can write 
\begin{align*}
f_0= 
\Bigl(\sqrt{1+\tilde\gamma^2}\Bigr)^{-1}-1-\lambda\Bigl(f^2\sqrt{1+\tilde\gamma^2}\Bigr)^{-1}.
\end{align*}
Clearly 
$\mp\lambda\parbb{f^2\sqrt{1+\tilde\gamma^2}}^{-1}
R(\lambda\mp\i0)\psi_\lambda^\pm[\xi]\in\vB_0^*$.

We write
 \begin{align*}
 1-\Bigl(\sqrt{1+\tilde\gamma^2}\Bigr)^{-1}= \dfrac{\tilde\gamma^2}{\sqrt{1+\tilde\gamma^2}\Bigl(1+\sqrt{1+\tilde\gamma^2}\Bigr)}.
 \end{align*}
 Thanks to
Proposition~\ref{prop:phase-space-radiation} we then conclude that for any $\psi\in C_{\mathrm c}^\infty(\R^d)$
\begin{align*}
\quad\hspace{-3mm}
\pm \left(\Bigl(\sqrt{1+\tilde\gamma^2}\Bigr)^{-1}-1\right)R(\lambda\mp\i0)\psi 
\in \vB_0^*.
\end{align*}
Now, for any $\varepsilon>0$ we can choose $\psi\in C_{\mathrm c}^\infty(\R^d)$ such that 
$\|\psi-\psi_\lambda^\pm[\xi]\|_\vB<\varepsilon$. 
Since the function $\Bigl(\sqrt{1+\tilde\gamma^2}\Bigr)^{-1}-1$ 
is bounded, 
we can then estimate
\begin{align*}
\Bigl\|\left(\Bigl(\sqrt{1+\tilde\gamma^2}\Bigr)^{-1}-1\right)
R(\lambda\mp\i0)(\psi-\psi_\lambda^\pm[\xi]) \Bigr\|_{\vB^*}
&\le 
C\|\psi-\psi_\lambda^\pm[\xi]\|_\vB<C\varepsilon.
\end{align*}
By combining these assertions we conclude that also
$$\pm\left(\Bigl(\sqrt{1+\tilde\gamma^2}\Bigr)^{-1}-1\right)
R(\lambda\mp\i0)\psi_\lambda^\pm[\xi]\in\vB_0^*.$$
Therefore also the second term on the right-hand side of
\eqref{eq:2010081635} belongs to 
$\vB_0^*$.

Hence we have 
\begin{equation*}
\chi^\perp(f<2)\bigl(f_{-2}p_f\pm1\bigr)\vF^\pm(\lambda)^*\xi 
\mp 2\phi_\lambda^\pm[\xi] \in\vB_0^*.
\end{equation*}
Then we can conclude that for any $\xi\in C_{\mathrm c}^\infty(\R^{d-1})$
\begin{equation}\label{eq:2010081740}
\xi=\omega_\pm\,\vSigmalim_{\rho\to \infty}
-\!\!\!\!\!\!\int_\rho \,\Bigl (J^{-1/2}\,\e^{\mp\i\theta_\lambda}
\Bigl(\tfrac12\pm \tfrac12 \parbb{f\sqrt{2r}}^{-1}p_f\Bigr)\vF^\pm(\lambda)^*\xi\Bigr)(f,\pm\cdot)\,\d f.
\end{equation}
By using Lemma~\ref{lem:H-boundedness} stated below and 
a continuity argument 
 we obtain that \eqref{eq:2010081740} holds for all $\xi\in\Sigma$, and
 therefore
\eqref{eq:aEigenfwB} holds.

The formula \eqref{eq:aEigenf2w} follows immediately from the first equality of \eqref{eq:aEigenfw} 
and Lemma~\ref{lem:uniquely-determine}. 
Thus we have (3).

By \eqref{eq:aEigenfw} and \eqref{eq:aEigenf2w} we have 
\begin{equation*}
c\|\xi\|_\Sigma 
\le \|\vF^\pm(\lambda)^*\xi\|_{\vB^*} 
\le C\|\xi\|_\Sigma. 
\end{equation*}
In particular $\vF^\pm(\lambda)^*$ is injective and has closed range. 
Therefore by Banach's closed range theorem \cite[Theorem P. 205]{Yo} 
the range of $\vF^\pm(\lambda)$ coincides with $\Sigma$, respectively. 
Hence we have shown (3) of Theorem\label{sec:reduct-radi-cond}~\ref{thm:char-gener-eigenf-1}. 
Moreover we can conclude that the range of 
$\delta(H-\lambda)=\vF^\pm(\lambda)^*\vF^\pm(\lambda)$ 
coincides with $\vE_\lambda$. 
Hence we have (4). 
\end{proof}

We note the following boundedness results.
\begin{lemma}\label{lem:H-boundedness} Let 
 $f_{-2}=\parbb{f\sqrt{f^2+g^2}}^{-1}$ and
 $\chi^\perp=\chi^\perp(f<\kappa)$ for any sufficiently big
 $\kappa\geq2$. Then
 \begin{subequations}
 \begin{align}\label{eq:relbnd}
&f_{-2}\chi^\perp p_f\text{ is }H\text{-bounded},\\&f_{-2}\chi^\perp
 p_f\vF^\pm(\lambda)^*\in\vL(\Sigma, \vB^*).\label{eq:relbnd2}
\end{align} 
 \end{subequations}
\end{lemma} 
\begin{proof}
As an operator on the set of smooth compactly supported function on
$\{r+x >2\}$ we calculate in terms of the operator $L$ from \eqref{eq:200818}
\begin{align*}
H_0=\tfrac12\parb{(p_f)^*\parb{f^2+g^2}^{-1}p_f-f^2}+\tfrac12 \parb{L+g^2},
\end{align*} and then in turn by a commutation formula
\begin{align*}
2f^{-1}H_0f^{-1}&=(p_f)^*f^2_{-2}p_f+f^{-1}\parb{L+g^2}f^{-1}+f_0;\\
 &f_0=f^{-4}\parb{f^2+g^2}^{-1}+\mathop{\mathrm{div}}\parb{ f^{-3} \nabla f}-1.
\end{align*} From Lemma \ref{lem:20092016} we see that $f_0$ is
bounded, and since the middle term is positive we can then bound
\begin{align*}
 &\chi^\perp (p_f)^*f^2_{-2}p_f\chi^\perp\\& \quad
 \quad \leq
2\Re\parb{H_0\parb{\chi^\perp f^{-1}}^2} +|\nabla \parb{\chi^\perp f^{-1}}|^2 -\chi^\perp f_0\chi^\perp \\& \quad
 \quad \leq 2\Re\parb{H_0\parb{\chi^\perp f^{-1}}^2}+C_1.
\end{align*} By \cs we can then bound
\begin{align*}
 \|{f_{-2}}p_f\chi^\perp \varphi\|^2\leq 2\|\varphi\|\, \|H_0\varphi\|+C_1 \| \varphi\|^2
 \text { for all }\varphi\in C_{\mathrm c}^\infty(\R^{d}).
\end{align*} Whence
\begin{align*}
 \|f_{-2}\chi^\perp p_f\varphi\|&\leq
 \|f_{-2}(\partial_f\chi^\perp) \varphi\|
 +C_2\|H_0\varphi\|+C_3\|\varphi\|\\\quad&\leq C_2\|H_0\varphi\|+C_4\|\varphi\|.
\end{align*} Since $C_{\mathrm c}^\infty(\R^{d})$ is a core of
$H_0$ and the graph-norms of $H_0$ and $H$ are equivalent,
\eqref{eq:relbnd} follows. 

For \eqref{eq:relbnd2} we need to bound, recalling the notation $\chi_m=\chi_m(f)=\chi(f/2^m<1)$,
\begin{align}\label{eq:bnfirst}
 \sup_{m\in\N,\,\norm{\xi}_\Sigma=1}\,2^{-m}\norm{f_{-2}\chi_m\chi^\perp
 p_f\vF^\pm(\lambda)^*\xi}_{\vH}^2<\infty.
\end{align} We estimate, more or less as above,
\begin{align*}
 &\chi^\perp \chi_m(p_f)^*f^2_{-2}p_f\chi_m\chi^\perp\\ \leq
2\Re\parb{(H-\lambda)\parb{\chi_m\chi^\perp f^{-1}}^2}& +|\nabla \parb{\chi_m\chi^\perp f^{-1}}|^2 +\parb{2(\lambda-q)f^{-2}-f_0}\parb{\chi_m\chi^\perp }^2.
 \end{align*}
 Taking
 expectation in
 $\chi_n\vF^\pm(\lambda)^*\xi\in\vD(H)$
 the first term
 to the right 
 does not
 contribute
 when taking
 $n\to \infty$. Since
 $\vF^\pm(\lambda)^*\in\vL(\Sigma,
 \vB^*)$, we 
 then 
 conclude (at
 least for any
 big 
 $\kappa$) 
 that 
\begin{align*}
 \sup_{m,\,\norm{\xi}=1}\,2^{-m}\norm{f_{-2}p_f\chi_m\chi^\perp
 \vF^\pm(\lambda)^*\xi}_{\vH}^2<\infty.
\end{align*} By a commutation, as in the proof of \eqref{eq:relbnd}, we finally deduce
\eqref{eq:bnfirst} and therefore also \eqref{eq:relbnd2}. 
 \end{proof}

\begin{remark*}
It is readily seen that one can remove the cut--off function $\chi^\perp(f<2)$ 
from \eqref{eq:relbnd}, however this stronger assertion is not needed in
the paper.
\end{remark*}

\section{Proof of radiation condition bounds}\label{sec:proof-prop-refprs}

Finally we prove Proposition~\ref{prop:phase-space-radiation}. 
The essence of the scheme of proof is given by the proof of Lemma~\ref{lem:200909}. 
We split the space into a classically forbidden region, a bounded
region and a classically allowed region.
The former regions can be treated in a rather straightforward fashion.
On the latter we compute the commutator
\begin{align*}
DP=\mathrm i[H,P]=\mathrm i(HP-PH)=2\mathop{\mathrm{Im}}(P(H-\lambda)),
\quad \lambda\in\mathbb R,
\end{align*}
replacing the classical time-derivative $D=\tfrac{\mathrm d}{\mathrm dt}$ by 
the Heisenberg derivative. 
Here $P$ is some moderated observable approximating $(f^s\gamma_\|)^2$ given later. 
More rigorously, we will first discuss a complex spectral parameter
and then take limit to the reals, 
so that we shall actually compute 
\begin{align}
\pm 2\mathop{\mathrm{Im}}(P(H-z)),\quad \pm\mathop{\mathrm{Im}}z>0
\label{eq:20090912}
\end{align}
rather than the above $DP$. 
The quantities \eqref{eq:20090912} should tend to be negative along
suitable scattering states 
due to what we have seen in the proof of Lemma~\ref{lem:200909}, or \eqref{eq:20090913},
and this would somehow verify the desired bound in quantum mechanics. 
We remark that our arguments employ only elementary techniques,
based entirely on the calculus of differential operators,
although the computations are very long.

\subsection{Preliminaries}

We first recall the LAP bound that we will often refer to. 
This was already proved in \cite{AIIS1}, 
and we do not discuss its proof in this paper. 
For a compact interval $I\subseteq \mathbb R$ set 
\begin{align*}
I_\pm=\{z=\lambda\pm\mathrm i\mu\in\mathbb C\,|\, \lambda\in I,\,\mu\in(0,1)\},
\end{align*}
respectively. 

\begin{thm}[{\cite[Theorem~2.8]{AIIS1}}]\label{thm:200909} 
Let $I\subseteq \mathbb R$ be a compact interval. 
Then there exists $C>0$ such that for any $\phi=R(z)\psi$
with $z\in I_\pm$ and $\psi\in C_{\mathrm c}^\infty(\mathbb R^d)$ 
 \begin{align*}
 &
\norm{\phi}_{\vB^*}
\le C\|\psi\|_{\mathcal B}
.
\end{align*} 
 \end{thm}

In regard to notation 
in what follows we shall simply write 
$$\theta=\theta_\lambda,\quad \gamma=\gamma(\lambda),\quad \gamma_\|=\gamma_\|(\lambda),$$ 
dropping the $\lambda$-dependence. 
There would not be any confusion. 
In addition, we continue to use the 
notation $\chi(\cdot <\kappa)$ and $\chi^\perp(\cdot<\kappa)$
from \eqref{eq:20100311} and \eqref{eq:20100312}, respectively. 
We shall also frequently use the following commutation
formulas in terms of any {$h\in C^\infty(\R^d)$,
  and with the repeated index summation convention,}
\begin{subequations}
\begin{align}\label{eq:ele1}
 p_ihp_i
&=
\mathop{\mathrm{Re}}(hp^2)
+\tfrac12 (\Delta h),\\
p_jp_ihp_ip_j
&=
p^2h p^2
-p_i(\partial_i\partial_jh) p_j
+p_i(\Delta h) p_i.\label{eq:ele3}
\end{align} 
\end{subequations}

\subsection{Classically forbidden region and bounded sets}\label{subsec:200913}

Here we treat $f$-bounded sets and a classically forbidden region. 
The following bounds are obviously sufficient for Proposition~\ref{prop:phase-space-radiation}
on the corresponding regions. 
We start with $f$-bounded sets.

\begin{lemma}\label{lem:200912}
Let $I\subseteq \mathbb R$ be a compact interval, 
and let {$\kappa>2$}.
Then for any $(k,\alpha)\in\mathbb N_0\times \mathbb N_0^d$ with $k+|\alpha|\le 2$
there exists $C>0$ such that 
for any $\phi=R(z)\psi$ with $z\in I_\pm$ and $\psi\in C_{\mathrm c}^\infty(\mathbb R^d)$ 
 \begin{align*}
 &
\norm{\chi(f<\kappa)|g|^kp^\alpha\phi}_{L^2}
\le C\|\psi\|_{\mathcal B}.
\end{align*} 
\end{lemma}
\begin{remark*} 
It is a routine exercise to check that any vector $\phi=R(z)\psi$
with $z\in \C\setminus\R$ and $\psi\in \vH_{-\infty}:=\cap_{k\in\N} \inp{r}^{-k}\vH$ obeys
\begin{align}\label{eq:reBND0} 
 \phi\in \vH^2_{-\infty}:=(p^2+1)^{-1}\vH_{-\infty},
\end{align} or equivalently, $ p^\alpha\phi\in \vH_{-\infty}$ for any $ \alpha\in \mathbb N_0^d$ with $|\alpha|\le 2$.
 Thanks to the
 second resolvent equation we can assume that $q_2=0$. Due to
 \eqref{eq:ele1} and \eqref{eq:ele3}
 \begin{align}\label{eq:reBND}
 \inp{r}^{-\abs{\alpha}/2}p^\alpha (H_0+q_1-z)^{-1}\in
 \vL(\vH)\text{ for }\abs{\alpha}\leq 2.
 \end{align} Now the bound \eqref{eq:reBND0} (with $q_2=0$) follows 
by
 repeated commutation of suitable powers of $\inp{r}$ and then repeated use
 of \eqref{eq:reBND}. We omit the details.
\end{remark*} 
\begin{proof}
In this proof let us write simply 
$$\eta=\chi(\sqrt{r}<\kappa/3),\quad 
\chi=\chi(f<\kappa),\quad 
\tilde\chi=\chi(f<3\kappa/2)
.$$
It suffices to prove the assertion only for large {$\kappa>2$}, 
so we may assume 
$$\mathop{\mathrm{supp}}q_2\subseteq \{\eta=1\}.$$

We first let $k=0$.
Then the bound is clear if $|\alpha|=0$ due to Theorem~\ref{thm:200909}. 
To treat the case $|\alpha|=1$ we use \eqref{eq:ele1}, Lemma~\ref{lem:20092016} and the bound 
\begin{align}\label{eq:ele2}
 x\leq x+r\leq f^2,
\end{align} allowing us to bound uniformly in $z\in I_\pm$ as 
\begin{align*}
 p\cdot \chi^2 p
\le 
2\mathop{\mathrm{Re}}\bigl(\chi^2 (H-z)\bigr)
-2q_2
+C_1\tilde\chi^2. 
\end{align*}
Then with the Cauchy-Schwarz inequality and Theorem~\ref{thm:200909}
\begin{align}
\begin{split}
\sum_{i=1}^d\|\chi p_i\phi\|_{L^2}^2
&
\le 
2\|\chi\phi\|_{\mathcal B^*}\|\psi\|_{\mathcal B}
+2\|\chi\phi\|_{L^2}\| q_2\phi\|_{L^2}
+C_1\|\tilde\chi \phi\|_{L^2}^2
\\&
\le 
C_2\|\psi\|_{\mathcal B}^2
+\|q_2\phi\|_{L^2}^2.
\end{split}
\label{eq:20091014}
\end{align}
Since $q_2$ is infinitesimally relatively $\Delta$-bounded, 
we have for any $\epsilon\in (0,1/2]$
\begin{align*}
\|q_2\phi\|_{L^2}
&
\le 
C_{3,\epsilon}\|\eta\phi\|_{L^2}+\epsilon\|(H-z-q_2)\eta\phi\|_{L^2}
\\
&\le 
C_{4,\epsilon}\|\psi\|_{\mathcal B}
+C_5\epsilon\sum_{i=1}^d\|\chi p_i\phi\|_{L^2}
+\epsilon\|q_2\phi\|_{L^2},
\end{align*}
so that 
\begin{align}
\|q_2\phi\|_{L^2}
&\le 
2C_{4,\epsilon}\|\psi\|_{\mathcal B}
+2C_5\epsilon\sum_{i=1}^d\|\chi p_i\phi\|_{L^2}.
\label{eq:20091015}
\end{align}
Here $C_{*,\epsilon}$ may be dependent on $\epsilon$, whereas $C_*$ are not. 
Now, by taking $\epsilon>0$ small enough, 
\eqref{eq:20091014}, \eqref{eq:20091015} and the Cauchy-Schwarz inequality 
imply the asserted bound with $k=0$ and $|\alpha|=1$. 
We can also conclude that 
\begin{align}\label{eq:qBND}
\|q_2\phi\|_{L^2}
\le C_6\|\psi\|_{\mathcal B}.
\end{align}
To discuss the case with $k=0$ and $|\alpha|=2$ 
we compute and estimate, using 
 \eqref{eq:ele1}, \eqref{eq:ele3}, \eqref{eq:ele2} and
 Lemma~\ref{lem:20092016},
\begin{align*}
p_jp_i\chi^2 p_ip_j
&=
p^2\chi^2 p^2
-p_i(\partial_i\partial_j\chi^2) p_j
+p_i(\Delta\chi^2) p_i
\\&\le 
4(H-z-q_2)^*\chi^2 (H-z-q_2)
+C_7p_i\tilde\chi^2 p_i
+C_7\tilde{\chi}^2.
\end{align*} 
Thus, invoking what we have proved above, we obtain 
\begin{align*}
\sum_{i,j=1}^d\|\chi p_ip_j\phi\|_{L^2}^2
&
\le 
4\|(H-z-q_2)\phi\|_{L^2}^2
+C_7\sum_{i=1}^d\|\tilde\chi p_i\phi\|_{L^2}^2
+C_7\|\chi\phi\|_{L^2}
\\&
\le 
C_8\|\psi\|_{\mathcal B}.
\end{align*}

Next we let $(k,\alpha)=(2,0)$. 
Note that for $f<\kappa$,
\begin{align}\label{eq:gBND}
g^2\le y^2\le -2\kappa^2x+\kappa^4.
\end{align}
Here the first inequality of \eqref{eq:gBND} is trivial,
while the second results from squaring $r<\kappa^2 -x$, 
which in turn follows from \eqref{eq:ele2} (and using
 that $f<\kappa$), and observing a cancellation of $x^2$. 
 By using \eqref{eq:gBND}, \eqref{eq:ele1} and
 Lemma~\ref{lem:20092016}, we can estimate
uniformly in $z\in I_\pm$ as 
\begin{align*}
\chi^2g^4
&\le 2\kappa^2\mathop{\mathrm{Re}}\bigl(\chi^2g^2(-x+q_1-z)\bigr)+C_9g^2\chi^2
\\&
\le 
2\kappa^2
\mathop{\mathrm{Re}}\bigl(\chi^2g^2(\tfrac12p^2-x+q_1-z)\bigr)
+ \tfrac12\kappa^2(\Delta \chi^2g^2)
+C_{9}\chi^2 g^2
\\&
\le 
2\kappa^2\mathop{\mathrm{Re}}\bigl(\chi^2g^2(H-z)\bigr)
-2\kappa^2g^2 q_2
+C_{10}\tilde\chi^2
+C_{9}\chi^2 g^2.
\end{align*}
This implies by the Cauchy-Schwarz inequality, \eqref{eq:qBND} and
Theorem~\ref{thm:200909} that 
\begin{align*}
\|\chi g^2\phi\|_{L^2}^2
&
\le 
2\kappa^2\|\chi g^2\phi\|_{L^2} \|\psi\|_{L^2}
+2\kappa^2\|\chi g^2\phi\|_{L^2} \|q_2\phi\|_{L^2}
\\&\phantom{{}={}}{}
+C_{10}\|\tilde\chi\phi\|_{L^2}^2
+C_{9}\|\chi g^2\phi\|_{L^2} \|\chi\phi\|_{L^2}
\\&
\le \tfrac12\|\chi g^2\phi\|_{L^2}^2
+C_{11}\|\psi\|_{\mathcal B}^2,
\end{align*}
finishing the case $(k,\alpha)=(2,0)$.

The bound for the case $(k,\alpha)=(1,0)$ follows from 
those for $(k,\alpha)=(0,0),(2,0)$ and the Cauchy-Schwarz inequality, 
or the case can be treated similarly as the case $(k,\alpha)=(2,0)$
 (the details are omitted). 
Finally we are left with $k=1$ and $|\alpha|=1$.
We bound (using again \eqref{eq:ele1}, \eqref{eq:ele2} and Lemma~\ref{lem:20092016})
\begin{align*}
p\cdot g^2\chi^2 p
&=
\mathop{\mathrm{Re}}(\chi^2 g^2p^2)
+\tfrac12(\Delta \chi^2g^2)\\ &
\le 
2\mathop{\mathrm{Re}}\bigl(\chi^2 g^2(H-z)\bigr)
-2g^2q_2
+C_{12}\parb{g^2\chi^2+\tilde\chi^2},
\end{align*}
and then we apply \eqref{eq:qBND} as well as the bounds for $(k,\alpha)=(0,0),(2,0)$ and the
Cauchy-Schwarz inequality.
\end{proof}

Next we discuss a classically forbidden region. 

\begin{lemma}\label{lem:200913}
Let $I\subseteq \mathbb R$ be a compact interval, 
and let $s> 0$, $\kappa>3$ and $\kappa'\in(0,1)$.
Then for any $(k,\alpha)\in\mathbb N_0\times \mathbb N_0^d$ with $k+|\alpha|\le 2$ there exists $C>0$ such that
for any $\phi=R(z)\psi$ with $z\in I_\pm$ and $\psi\in C_{\mathrm c}^\infty(\mathbb R^d)$ 
 \begin{align*}
 &
\norm{\chi^\perp(f<\kappa)\chi(f/|g|<\kappa')f^{s+1}|g|^kp^\alpha\phi}_{L^2_{-1/2}}
\le C\|f^s\psi\|_{L^2_{1/2}}.
\end{align*}
\end{lemma}
\begin{remarks*}
\begin{enumerate}[1)]
\item
For sufficiently large $\kappa>3$ the above
 $\chi$-factors 
 may be considered 
as a localization to a `classically forbidden region' in position
space. This is immediate from the bound \eqref{eq:clasFORB} given below 
(for orbits of energy $\lambda\in I$).
\item
We may replace the weight $|g|^k$ by $f^k$ 
on the left-hand side of the asserted bound due to the fact that $f< \kappa'|g|$ 
on $\mathop{\mathrm{supp}}\chi(f/|g|<\kappa')$. This is certainly a weaker statement,
but we will use it later as well.
\end{enumerate}
\end{remarks*}
\begin{proof}
In this proof let us abbreviate 
 $\chi=\chi^\perp(f<\kappa)\chi(f/|g|<\kappa')$ and choose a
 similar function $\tilde\chi$, say
 $\tilde\chi=\chi^\perp(f<2\kappa/3)\chi(f/|g|<\tilde{\kappa})$,
 where $\tilde{ \kappa}\in (\kappa',1)$ and
 $\chi(\cdot<\tilde{\kappa})$ agrees with \eqref{eq:20100311} but is chosen `sufficiently sharply
 localized' such that $\tilde\chi=1$ on the support of $\chi$. (This
 extension procedure will be iterated finitely many times below.)
By Lemma~\ref{lem:200912} we may assume $\kappa>3$ is sufficiently large so that 
$$q_2=0\ \ \text{on }\mathop{\mathrm{supp}}\tilde\chi.$$
Due to $1\le f\le \kappa'g$ it suffices to consider the case $k=2-|\alpha|$.

First we let $\alpha=0$. 
Since $f^2\le \kappa'^2g^2$ and $\kappa>3$ is large, we may bound 
{for some $c_1,C_1>0$
\begin{align*}
 g^2-f^2 +2q_1-2\mathop{\mathrm{Re}}(z)\geq 
g^2-f^2-C_1
\geq \tfrac12(1-\kappa'^2) g^2\geq c_1g^2,
\end{align*} and therefore with $C_2=2c_1^{-1}$ 
\begin{align}\label{eq:clasFORB}
 g^2\le C_2\mathop{\mathrm{Re}}(-x+q_1-z).
\end{align}
}This is uniform in $z\in I_\pm$, and we can combine it with
 \eqref{eq:ele1} and Lemma~\ref{lem:20092016} to bound
\begin{align*}
\chi^2f^{2s+1}g^4
&
\le 
C_2\mathop{\mathrm{Re}}\bigl(\chi^2f^{2s+1}g^2(\tfrac12p^2-x+q_1-z)\bigr)
+\tfrac14C_2(\Delta \chi^2f^{2s+1}g^2)
\\&
\le 
C_2\mathop{\mathrm{Re}}\bigl(\chi^2f^{2s+1}g^2(H-z)\bigr)
+C_3\tilde\chi^2f^{2s+1}.
\end{align*}
This implies by the Cauchy-Schwarz inequality
\begin{align*}
\|\chi f^{s+1}g^2\phi\|_{L^2_{-1/2}}^2
&
\le 
C_4\|\chi f^{s+1}g^2\phi\|_{L^2_{-1/2}} \|f^s\psi\|_{L^2_{1/2}}
+C_4\|\tilde\chi f^{s+1}\phi\|_{L^2_{-1/2}}^2
,
\end{align*}
or 
\begin{align*}
\|\chi f^{s+1}g^2\phi\|_{L^2_{-1/2}}^2
&
\le 
C_5\|f^s\psi\|_{L^2_{1/2}}^2
+C_5\|\tilde\chi f^{s+1}\phi\|_{L^2_{-1/2}}^2
.
\end{align*}
Thus we have reduced the weight from $f^{s+1}g^2$ to $f^{s+1}\le (\kappa')^2f^{s-1}g^2$
at the expense of replacing $\chi$ by the slightly wider cut-off function $\tilde\chi$. 
Letting $\kappa>3$ be sufficiently large from the beginning and repeating the above reduction 
until a negative exponent shows up, 
we can verify the asserted bound for $\alpha=0$ by Theorem~\ref{thm:200909}. 

As for the first term with $|\alpha|=1,2$, 
we can compute and estimate similarly as in the proof of Lemma~\ref{lem:200912}.
We have 
\begin{align}\label{eq:first}
\begin{split}
p\cdot f^{2s+1}g^2&\chi^2 p
=
\mathop{\mathrm{Re}}(\chi^2 f^{2s+1}g^2p^2)
+\tfrac12(\Delta \chi^2f^{2s+1}g^2)
\\&
\le 
2\mathop{\mathrm{Re}}\bigl(\chi^2 f^{2s+1}g^2(H-z)\bigr)
+C_6\tilde\chi^2f^{2s+1},
\end{split}
\end{align}
and 
\begin{align*}
 \begin{split}
 p_jp_i\chi^2 f^{2s+1} &p_ip_j
=
p^2\chi^2 f^{2s+1}p^2
-p_i(\partial_i\partial_jf^{2s+1}\chi^2) p_j
+p_i(\Delta f^{2s+1}\chi^2) p_i
\\&\le 
4(H-z)^*\chi^2 f^{2s+1}(H-z)
+C_7p_i\tilde\chi^2 f^{2s+1}p_i
+C_7\tilde\chi^2f^{2s+1}. 
 \end{split}
\end{align*}
Then the asserted bounds for $\abs{\alpha}=1,2$ 
are almost trivial, and 
 we are done.
\end{proof}

\subsection{Classically allowed region}

\subsubsection{The operator $P$ and preliminary lemmas}

Finally we discuss a classically allowed region. 
We compute and bound the quantity \eqref{eq:20090912} for the operator $P$
of the form
\begin{subequations}
\begin{align}\label{eq:good1}
 P=P_{\kappa,\kappa',\nu}^s=\gamma_\|\chi^2\Theta^{2s}\gamma_\|,\quad s\in(0,1+\delta).
\end{align}
The radiation operator $\gamma_\|=\gamma_\|(\lambda)$ is from \eqref{eq:200820}. 
The function $\chi$ is given by 
\begin{align}\label{eq:good2}
\chi=\chi^\perp(f<\kappa)\chi^\perp(f/|g|<\kappa'),\quad 
\kappa>3,\ \ \kappa'\in (0,1/2).
\end{align}
This is a smooth cut-off function complementing those in Subsection~\ref{subsec:200913}.
The same short notation $\chi$ was adopted for different cut-off functions in the proofs 
of Lemmas~\ref{lem:200912} and \ref{lem:200913},
but in this subsection we use $\chi$ only in the above sense. 
The function $\Theta$ is a regularized weight defined as 
\begin{align}\label{eq:good3}
\Theta=\Theta_{\nu}
=\int_0^{f}(1+s/2^\nu)^{-2}\,\mathrm ds,\quad \nu\in \mathbb N_0.
\end{align} 
\end{subequations} 
This is a bounded function for each $\nu\in\mathbb N_0$, 
and approaches $f$ from below as $\nu\to\infty$. 
We record some elementary bounds of its derivatives as a lemma. 
We denote the derivatives viewed as a function of $f$ 
simply by $\Theta',\Theta'',\dots,\Theta^{(k)}$. 

\begin{lemma}\label{lem:basic30} 
One has uniformly in $\nu\in\N_0$ and $k\in \N$
\begin{align*} 
&\tfrac12 \le \Theta\le \min\{2^\nu,f \}
,\quad 
0\le
(-1)^{k-1}\Theta^{(k)}\le
k!f^{-k}\Theta.
\end{align*}
\end{lemma}
\begin{proof}
These bounds are elementary, and we omit the proof. 
\end{proof}

The following identity is a quantum counterpart of \eqref{eq:quad1}
for energy $\lambda\in\mathbb R$
and plays an essential role in quantum mechanics 
as \eqref{eq:quad1} did in the classical context. 

\begin{lemma}\label{eq:200807b}
Let $\kappa>3$ be sufficiently large (depending on the support of $q_2$). 
Then on the open set $\set{f>2\kappa/3}$ one has 
for any $\lambda\in\mathbb R$ 
\begin{align*}
\begin{split}
H-\lambda
&=
\tfrac12\gamma^2
\pm\tfrac12r^{-1}(f^2+\lambda)\gamma_\|
\mp\tfrac{\mathrm i}4fr^{-2}(g^2-\lambda)
+\tfrac14r^{-1}(g^2-\lambda)^2
+q_1
,
\end{split}
\end{align*}
or equivalently 
\begin{align*}
\begin{split}
\gamma_\|
&=
\pm 2r(f^2+\lambda)^{-1}\bigl(H-\lambda
-\tfrac12\gamma^2
\pm\tfrac{\mathrm i}4fr^{-2}(g^2-\lambda)
-\tfrac14r^{-1}(g^2-\lambda)^2
-q_1\bigr)
.
\end{split}
\end{align*}
\end{lemma}
\begin{proof}
By definitions \eqref{eq:QM1} and \eqref{eq:Jac1}
we can write, letting $\gamma=\gamma(\lambda)$ and $\theta=\theta_\lambda$, 
\begin{align*}
H-\lambda
&=
\tfrac12\gamma^2
\pm\mathop{\mathrm{Re}}\bigl((\nabla\theta)\cdot\gamma\bigr)
+\tfrac12|\nabla\theta|^2
-x+q_1-\lambda
.
\end{align*}
By Lemma~\ref{lem:200821} we can further compute the second and third terms on the above right-hand side as 
\begin{align*}
\mathop{\mathrm{Re}}\bigl((\nabla\theta)\cdot\gamma\bigr)
&=
\mathop{\mathrm{Re}}\bigl((f^2+\lambda)(\nabla f)\cdot\gamma\bigr)
\\&=
r^{-1}(f^2+\lambda)\mathop{\mathrm{Re}}\bigl(r(\nabla f)\cdot\gamma\bigr)
-\tfrac{\mathrm i}2r(\nabla f)\cdot \bigl(\nabla(f^2+\lambda)r^{-1}\bigr)
\\&=
\tfrac12r^{-1}(f^2+\lambda)\gamma_\|
-\tfrac{\mathrm i}4fr^{-2}(g^2-\lambda)
,
\end{align*}
and 
\begin{align*}
\tfrac12|\nabla\theta|^2
=
\tfrac12(f^2+\lambda)^2|\nabla f|^2
=
\tfrac14f^4r^{-1}+\tfrac12\lambda f^2r^{-1}+\tfrac14\lambda^2r^{-1},
\end{align*}
respectively. From the above identities we obtain the assertion.
\end{proof}

Let us present some more lemmas for the later arguments. 

\begin{lemma}\label{lem:200902}
On the open set $\set{r+x>2}$ one has 
\begin{align*}
\mathrm i[\gamma^2,\gamma_\|]
&=
2\gamma \cdot h\gamma
-\tfrac12(\Delta fr^{-1})-\tfrac{d-2}2(\Delta f^{-1}),
\end{align*}
where $h$ is from \eqref{eq:20090623}.
\end{lemma}
\begin{proof}
By \eqref{eq:200820} we compute it as 
\begin{align}
\begin{split}
\mathrm i[\gamma^2,\gamma_\|]
&=
-2\mathop{\mathrm{Im}}(\gamma^2\gamma_\|)
\\&
=
-4\mathop{\mathrm{Im}}(\gamma^2r(\nabla f)\cdot \gamma)
+\mathop{\mathrm{Re}}(\gamma^2fr^{-1})
+(d-2)\mathop{\mathrm{Re}}(\gamma^2f^{-1}).
\end{split}
\label{eq:200827}
\end{align}
As for the first term of \eqref{eq:200827}, noting that $\gamma_j$ and $\gamma_k$ commute, 
we have 
\begin{align*}
-4\mathop{\mathrm{Im}}(\gamma^2r(\nabla f)\cdot \gamma)
&=
-4\mathop{\mathrm{Im}}(\gamma_j\gamma_jr(\partial_k f)\gamma_k)
\\&=
4\mathop{\mathrm{Re}}(\gamma_j(\partial_jr\partial_k f)\gamma_k)
-4\mathop{\mathrm{Im}}(\gamma_jr(\partial_k f)\gamma_k\gamma_j)
\\&=
2\gamma \cdot h\gamma
-\gamma \cdot fr^{-1}\gamma
-(d-2)\gamma \cdot f^{-1}\gamma.
\end{align*}
The remaining part of \eqref{eq:200827} is computed as follows using
\eqref{eq:ele1}, 
\begin{align*}
&\mathop{\mathrm{Re}}(\gamma^2fr^{-1})
+(d-2)\mathop{\mathrm{Re}}(\gamma^2f^{-1})
\\&
=
\gamma \cdot fr^{-1}\gamma
-\tfrac12(\Delta fr^{-1})
+(d-2)\gamma\cdot f^{-1}\gamma
-\tfrac{d-2}2(\Delta f^{-1}).
\end{align*}
Thus we obtain the assertion.
\end{proof}

We will use the term $\gamma^2\chi^2f^{-1}\Theta^{2s}\gamma^2$ as one of the `leading terms', 
and the following `ellipticity bounds' will come in handy. 
These bounds may
 be seen as 
 easy consequences of \eqref{eq:ele1} and \eqref{eq:ele3} (by a
 conjugation by a phase factor), and as
 before we use 
Einstein's convention. 
Let us set 
\begin{align}
\eta=\chi^\perp(f<2\kappa/3)\chi^\perp(f/|g|<2\kappa'/3)-\chi^\perp(f<3\kappa/2)\chi^\perp(f/|g|<3\kappa'/2),
\label{eq:200918}
\end{align}
which is designed to be supported in the union of a
 bounded set and a
 classically forbidden region. 
 Note also that $\eta=1$ on
 $\mathop{\mathrm{supp}}(\nabla\chi)$. 
\begin{lemma}
 \label{lem:algebra0} 
Let $s>0$, $\kappa>3$, $\kappa'\in(0,1/2)$ and $t\in\mathbb R$.
Then there exists $C>0$ such that for any $\epsilon\in (0,1)$ and $\nu\in\mathbb N_0$
\begin{subequations}
 \begin{align}\label{eq:1ell}
 \gamma_i \chi^2f^t\Theta^{2s}\gamma_i
\leq \epsilon \gamma^2
 \chi^2f^{-1}\Theta^{2s}\gamma^2
+C\epsilon^{-1}\chi^2 f^{2t+1}\Theta^{2s}
+Cf^{t-4}\Theta^{2s}
 \end{align}
 and 
 \begin{align}\label{eq:2ell}
\pm T
\leq 
\epsilon\gamma^2\chi^2f^{-1}\Theta^{2s}\gamma^2 
+\gamma^2\eta^2f^{-1}\Theta^{2s}\gamma^2 
+C\epsilon^{-1}f^{-9}\Theta^{2s},
 \end{align} where $T=\gamma_j\gamma_i
 \chi^2f^{-1}\Theta^{2s}\gamma_i\gamma_j-\gamma^2\chi^2f^{-1}\Theta^{2s}\gamma^2
 .$ 
\end{subequations}
\end{lemma}
\begin{proof}
We estimate using \eqref{eq:ele1}, the Cauchy-Schwarz inequality,
Lemma~\ref{lem:basic30} and Lemma~\ref{lem:20092016}
\begin{align*}
\gamma_j\chi^2f^t\Theta^{2s}\gamma_j&
=\mathop{\mathrm{Re}}\bigl(\chi^2f^t\Theta^{2s}\gamma^2\bigr)+\tfrac12(\Delta \chi^2f^t\Theta^{2s})\\
&\leq \epsilon \gamma^2\chi^2f^{-1}\Theta^{2s}\gamma^2+\tfrac1{4\epsilon}\chi^2f^{2t+1}\Theta^{2s}
+C_1f^{t-4}\Theta^{2s}.
\end{align*} 
Here we used that on
 $\mathop{\mathrm{supp}}\partial_i\chi$ either $r$ is bounded or $\abs{g}\simeq f\simeq \sqrt{r}$. 
This proves \eqref{eq:1ell}.
Noting that $\gamma_i$ and $\gamma_j$ commute we have, cf. \eqref{eq:ele3}, 
\begin{align*}
 \gamma_j\gamma_i\chi^2f^t\Theta^{2s}\gamma_i\gamma_j
= \gamma^2\chi^2f^t\Theta^{2s} \gamma^2
-\gamma_j(\nabla^2_{ij}\chi^2f^t\Theta^{2s} )\gamma_i
+\gamma_j(\Delta \chi^2f^t\Theta^{2s})\gamma_j.
\end{align*} 
We take here $t=-1$ and note that \eqref{eq:1ell} applies to
 $\chi^2$ (as stated) as well as to $\chi^2$ replaced by $\eta^2$,
 in particular in both cases with $t=-5$. The last two terms are
bounded in this way,
and we obtain \eqref{eq:2ell}. 
\end{proof}

\subsubsection{Commutator estimates}

Now we are ready to bound \eqref{eq:20090912}. 
To prove Proposition~\ref{prop:phase-space-radiation} 
we need to bound \eqref{eq:20090912} from both above and below.
 We start with a bound from above which 
technically, although elementary, is the most demanding issue of the
paper. To facilitate the upcoming (non-trivial) bookkeeping task it is
convenient to introduce (in terms of the function $\eta$ from
\eqref{eq:200918} depending on parameters $\kappa$ and $\kappa'$) 
the form
\begin{align}\label{eq:bookk}
 Q_s=p_jp_i\eta^2\Theta^{2s}p_ip_j+p_j\eta^2f^2\Theta^{2s}p_j
 +\eta^2f^4\Theta^{2s};\quad s>0.
\end{align} 
\begin{lemma}\label{lem:200919}
Let $I\subseteq \mathbb R$ be a compact interval, 
$\kappa>3$ large enough, and $\kappa'\in (0,1/2)$. 
For any $s\in (0,2)$ 
there exist $c,C>0$ such that 
uniformly in $\nu\in\mathbb N_0$ and $z=\lambda\pm \i \mu\in I_\pm$, 
as a form on $\vH^2_{-\infty}$ (see \eqref{eq:reBND0}), 
\begin{align*} 
\pm\mathop{\mathrm{Im}}(P(H-z))
&
\le 
-c\gamma_\|\chi^2 f^{-1}\Theta^{2s}\gamma_\|
+{C\mu f^{-2}\Theta^{2s}}
\\&\phantom{{}={}}{}
+Cf^{-3-2\delta}\Theta^{2s}+C(H-z)^*f\Theta^{2s}(H-z) +CQ_s .
\end{align*} 
\end{lemma}
\begin{remark*}
This is clearly a quantum counterpart of \eqref{eq:20090913}. 
In Subsection \ref{subsec:propPr} we take its expectation in the state $\phi=R(z)\psi$. 
The 
 second term on the right-hand side disappears in the $\mu\to 0$ limit, 
the third term can be bounded by Theorem~\ref{thm:200909}
for $s\in (0,1+\delta)$, 
 the fourth term is directly bounded by a weighted norm of $\psi$ and 
the fifth term by Lemmas~\ref{lem:200912} and \ref{lem:200913}. 
We
obtain in this way an essentially negative upper bound of the
expectation of the the left-hand side of the estimate in
$\phi=R(\lambda\pm\i 0)\psi$ which is uniform in $\nu$, allowing us
finally to take $\nu\to \infty$ (the contribution from the left-hand side of the estimate 
can be regarded as
small due to the factor $H-z$).
In conclusion we obtain the part of
Proposition~\ref{prop:phase-space-radiation} involving
$\gamma_\|$ for the classical allowed
region. Furthermore Subsection 
\ref{subsec:200913} allows us to remove the cut-off function
$\chi$. The desired bounds for $\gamma$ and $\tilde\gamma$ show up in our
scheme of proof parallelly to the classical mechanics case, 
now employing Lemma~\ref{eq:200807b}. 
\end{remark*}
\begin{proof}
\emph{Step I}.
For simplicity let us treat only the upper sign, recalling 
\eqref{eq:QM1}, \eqref{eq:200820} and \eqref{eq:good1}--\eqref{eq:good3}.
The following estimates are uniform in $z=\lambda+\mathrm i\mu\in I_+$. 
By Lemma~\ref{eq:200807b} with upper sign we first rewrite 
\begin{align}
\begin{split}
&\mathop{\mathrm{Im}}(P(H-z))
\\&
=
\tfrac12\mathop{\mathrm{Im}}\bigl(\gamma_\| \chi^2\Theta^{2s}\gamma_\| \gamma^2\bigr)
+\tfrac12\mathop{\mathrm{Im}}\bigl(\gamma_\| \chi^2\Theta^{2s}\gamma_\|r^{-1}(f^2+\lambda)\gamma_\|\bigr)
\\&\phantom{{}={}}{}
-\tfrac14\mathop{\mathrm{Re}}\bigl(\gamma_\| \chi^2\Theta^{2s}\gamma_\| fr^{-2}(g^2-\lambda)\bigr)
+\tfrac14\mathop{\mathrm{Im}}\bigl(\gamma_\| \chi^2\Theta^{2s}\gamma_\|r^{-1}(g^2-\lambda)^2\bigr)
\\&\phantom{{}={}}{}
+\mathop{\mathrm{Im}}\bigl(\gamma_\| \chi^2\Theta^{2s}\gamma_\|q_1\bigr)
-\mu \gamma_\| \chi^2\Theta^{2s}\gamma_\|. 
\end{split}
\label{2008078b}
\end{align}
We are going to further compute and bound the right-hand side of \eqref{2008078b}. 
Due to the cut-off function $\chi$ we have on its support
$$\langle g\rangle=(1+g^2)^{1/2}\le C_1f,\quad f^2\le 2 r\le C_1f^2.$$
In
 particular these bounds allow use in combination with
 Lemmas~\ref{lem:20092016} and \ref{lem:basic30} to record the
 following bounds to be used tacitly below,
\begin{align*}
 \partial^\alpha \chi=\vO\parb{f^{-2\abs{\alpha}}},\quad \partial^\alpha \Theta=\vO\parb{f^{-2\abs{\alpha}}}\Theta,\quad\partial^\alpha \theta_\lambda=\vO\parb{f^{3-2\abs{\alpha}}},\quad \partial^\alpha F_j=\vO\parb{f^{1-2\abs{\alpha}}};
\end{align*} here 
the components of vector field of \eqref{eq:200820} are denoted $F_j=2r\partial_j f$.
We shall gather and write {the third term} to the last term on the
right-hand side of the bound of the lemma (up to a constant) as 
\begin{align*}
Q
&=
f^{-3-2\delta}\Theta^{2s}
+(H-z)^*f\Theta^{2s}(H-z) +Q_s
\end{align*}
for short. 
Given our goal a term bounded by $Q$ may be called an 
\emph{admissible error term}.
{The second term $\mu f^{-2}\Theta^{2s}$ of the bound may 
also be called {admissible}, but rather we separate it out for a discussion in Remark~\ref{rem:20:1030}.}
Note for example that 
for any real $f_\c\in C^\infty_{\c}(\R^d)$ 
whose support does not intersect $\mathop{\mathrm{supp}}q_2$ the operator
$\gamma_\|f_\c\Theta^{2s}\gamma_\|$ is admissible, i.e.\ there exists
$C_2>0$ (independent of $\nu$ and $z$) such that
\begin{subequations}
\begin{align}
 \label{eq:Fbnd} -C_2 Q\leq \gamma_\|f_\c\Theta^{2s}\gamma_\|\leq C_2 Q.
\end{align} 
This can be verified e.g.\ by a estimation similar to \eqref{eq:first} and the 
Cauchy-Schwarz inequality. Another example we will need is
\begin{align}\label{eq:TT1}
 T^*\Theta^{2s}T\leq C_2'Q; \quad T=f(\nabla \chi)\cdot \gamma \gamma_\|.
\end{align} 
Indeed this follows readily by first writing $T$ as a finite sum of
terms of the form $\vO(f^0)p_ip_j+\vO(f)p_j +\vO(f^2)$ and then
using \cS. 
 Similarly 
\begin{align}\label{eq:TT2}
 T_j^*\Theta^{2s}T_j\leq C''_{2}Q; \quad T_j=f^{2}(\partial_j
 \chi)\gamma_\|.
\end{align} 
\end{subequations}

\noindent
\emph{Step II}.
Now we compute each term on the right-hand side of \eqref{2008078b}.
As for the first term of \eqref{2008078b}, 
by Lemma
 \ref{lem:200902}, \eqref{eq:Fbnd}, \eqref{eq:TT1}
and (repeated applications of) the Cauchy-Schwarz inequality
\begin{align}\label{eq:1first}
\begin{split}
\tfrac12\mathop{\mathrm{Im}}\bigl(\gamma_\| \chi^2\Theta^{2s}\gamma_\| \gamma^2\bigr)
&
=
\tfrac12\mathop{\mathrm{Re}}\bigl(\gamma_\| \chi^2\Theta^{2s}\mathrm i[\gamma^2,\gamma_\|]\bigr)
+\tfrac12\mathop{\mathrm{Re}}\bigl(\gamma_\| (\nabla\chi^2\Theta^{2s})\cdot\gamma \gamma_\| \bigr)
\\&
\le 
\mathop{\mathrm{Re}}\bigl(\gamma_\| \chi^2\Theta^{2s}\gamma\cdot h\gamma\bigr)
+\tfrac s2\mathop{\mathrm{Re}}\bigl(\gamma_\| \chi^2r^{-1}\Theta'\Theta^{2s-1}\gamma_\|^2\bigr)
\\&\phantom{{}={}}{}
+\epsilon \gamma_\|\chi^2fr^{-1}\Theta^{2s}\gamma_\|
+C_{3,\epsilon}Q. 
\end{split}
\end{align}
Here $\epsilon\in (0,1)$ is a small parameter fixed in the last step of the proof.
We remark, here and below, that, if $\nabla$ acts on $\chi^2$, 
the corresponding term can be partly absorbed 
into $Q$ by the Cauchy-Schwarz inequality.

The estimates for the other terms of \eqref{2008078b} are basically the same. 
As for the second term of \eqref{2008078b}, 
by Lemmas~\ref{lem:20092016} 
and \ref{lem:basic30}, \eqref{eq:Fbnd}, 
 \eqref{eq:TT2} and 
the Cauchy-Schwarz inequality
\begin{align}
\begin{split}
&
\tfrac12\mathop{\mathrm{Im}}\bigl(\gamma_\| \chi^2\Theta^{2s}\gamma_\|r^{-1}(f^2+\lambda)\gamma_\|\bigr)
\\&
=
\tfrac12\gamma_\| 
\Bigl((f^2+\lambda)(\nabla f)\cdot(\nabla \chi^2\Theta^{2s})
-\chi^2\Theta^{2s}r(\nabla f)\cdot(\nabla r^{-1}(f^2+\lambda))\Bigr)\gamma_\|
\\&
\le 
\tfrac s2\gamma_\| \chi^2 fr^{-1}\Theta^{2s}\gamma_\|
-\tfrac14\gamma_\| \chi^2fg^2r^{-2}\Theta^{2s}\gamma_\|
+\epsilon \gamma_\|\chi^2fr^{-1}\Theta^{2s}\gamma_\|
+{C_{4,\epsilon}Q}
.
\end{split}
\end{align}
As for the third term of \eqref{2008078b}, 
similarly to the above 
\begin{align}
\begin{split}
&
-\tfrac14\mathop{\mathrm{Re}}\bigl(\gamma_\| \chi^2\Theta^{2s}\gamma_\| fr^{-2}(g^2-\lambda)\bigr)
\\&
= 
-\tfrac14\gamma_\| \chi^2\Theta^{2s} fr^{-2}(g^2-\lambda)\gamma_\|
-\tfrac12\mathop{\mathrm{Im}}\bigl(\gamma_\| \chi^2\Theta^{2s}r(\nabla f)\cdot (\nabla fr^{-2}(g^2-\lambda))\bigr)
\\&
\le 
-\tfrac14\gamma_\| \chi^2 fg^2r^{-2}\Theta^{2s}\gamma_\|
+\epsilon \gamma_\|\chi^2fr^{-1}\Theta^{2s}\gamma_\|
+\epsilon\chi^2f^{-1}r^{-2}g^8\Theta^{2s} 
+C_{5,\epsilon}Q.
\end{split}
\end{align}
As for the fourth term of \eqref{2008078b}, 
similarly to the above 
\begin{align}
\begin{split}
&\tfrac14\mathop{\mathrm{Im}}\bigl(\gamma_\| \chi^2\Theta^{2s}\gamma_\| r^{-1}(g^2-\lambda)^2\bigr)
\\&
=
-\tfrac12\mathop{\mathrm{Re}}\bigl(\gamma_\| \chi^2\Theta^{2s}r(\nabla f)\cdot(\nabla r^{-1}(g^2-\lambda)^2)\bigr)
\\&
\le 
\tfrac14\mathop{\mathrm{Re}}\bigl(\gamma_\| \chi^2fg^4r^{-2}\Theta^{2s}\bigr)
+\epsilon \gamma_\|\chi^2fr^{-1}\Theta^{2s}\gamma_\|
+\epsilon\chi^2f^{-1}r^{-2}g^8\Theta^{2s}
+C_{6,\epsilon}Q.
\end{split}
\end{align}
As for the fifth term of \eqref{2008078b}, similarly to the above 
\begin{align}
\begin{split}
\mathop{\mathrm{Im}}\bigl(\gamma_\| \chi^2\Theta^{2s}\gamma_\|q_1\bigr)
&=
-2\mathop{\mathrm{Re}}\bigl(\gamma_\| \chi^2\Theta^{2s}r(\nabla f)\cdot(\nabla q_1)\bigr)
\\&
\le 
\epsilon \gamma_\|\chi^2fr^{-1}\Theta^{2s}\gamma_\|
+C_{7,\epsilon}Q.
\end{split}
\end{align}
The last term of \eqref{2008078b} is obviously bounded as 
\begin{align}
-\mu\gamma_\| \chi^2\Theta^{2s}\gamma_\|\le 0. 
\label{eq:200917}
\end{align}
Hence by \eqref{eq:1first}--\eqref{eq:200917} we obtain 
\begin{align}
\begin{split}
\mathop{\mathrm{Im}}(P(H-z))
&
\le 
\mathop{\mathrm{Re}}\bigl(\gamma_\| \chi^2\Theta^{2s}\gamma\cdot h\gamma\bigr)
+\tfrac s2\mathop{\mathrm{Re}}\bigl(\gamma_\| \chi^2r^{-1}\Theta'\Theta^{2s-1}\gamma_\|^2\bigr)
\\&\phantom{{}={}}{}
+\tfrac14\mathop{\mathrm{Re}}\bigl(\gamma_\| \chi^2fg^4r^{-2}\Theta^{2s}\bigr)
-\tfrac12\gamma_\| \chi^2fg^2r^{-2}\Theta^{2s}\gamma_\|
\\&\phantom{{}={}}{}
+(\tfrac s2+5\epsilon)\gamma_\| \chi^2 fr^{-1}\Theta^{2s}\gamma_\|
+2\epsilon\chi^2f^{-1}r^{-2}g^8\Theta^{2s}
+C_{8,\epsilon}Q
. 
\end{split}
\label{eq:20091716}
\end{align}

\smallskip
\noindent
\emph{Step III}.
Next we compute and bound the first to third terms on the right-hand side of \eqref{eq:20091716}.
These are odd order operators, and do not seem to have sign, 
but they actually give a good negative contribution by using Lemma~\ref{eq:200807b}
along with the fourth term of \eqref{eq:20091716}.
As for the first term of \eqref{eq:20091716},
we first rewrite it by substituting the adjoint of the second expression of Lemma~\ref{eq:200807b} as 
\begin{align*}
&\mathop{\mathrm{Re}}\bigl(\gamma_\| \chi^2\Theta^{2s}\gamma\cdot h\gamma\bigr)
\\&
=
2\mathop{\mathrm{Re}}\bigl((H-z)^*r(f^2+\lambda)^{-1}\chi^2\Theta^{2s}\gamma\cdot h\gamma\bigr)
+2\mu \mathop{\mathrm{Im}}\bigl(\chi^2r(f^2+\lambda)^{-1}\Theta^{2s}\gamma\cdot h\gamma\bigr)
\\&\phantom{{}={}}{}
-\mathop{\mathrm{Re}}\bigl(\gamma^2\chi^2r(f^2+\lambda)^{-1}\Theta^{2s}\gamma\cdot h\gamma\bigr)
+\tfrac12\mathop{\mathrm{Im}}\bigl(\chi^2fr^{-1}(f^2+\lambda)^{-1}(g^2-\lambda)\Theta^{2s}\gamma\cdot h\gamma\bigr)
\\&\phantom{{}={}}{}
-\tfrac12\mathop{\mathrm{Re}}\bigl(\chi^2(f^2+\lambda)^{-1}(g^2-\lambda)^2\Theta^{2s}\gamma\cdot h\gamma\bigr)
-2\mathop{\mathrm{Re}}\bigl(\chi^2r(f^2+\lambda)^{-1}q\Theta^{2s}\gamma\cdot h\gamma\bigr)
.
\end{align*}
The right-hand side of the above identity is estimated by using 
Lemmas~\ref{lem:20092016}, \ref{lem:basic30} and \ref{lem:algebra0}, 
\eqref{eq:20090623} and the Cauchy-Schwarz inequality. 
This is long but based on elementary computations, 
and we do not present all the details. 
Only the second term requires a slightly special treatment,
so let us discuss it. 
By the Cauchy-Schwarz inequality and Lemma~\ref{lem:algebra0} 
we can bound it as 
\begin{align*}
2\mu \mathop{\mathrm{Im}}\bigl(\chi^2r(f^2+\lambda)^{-1}\Theta^{2s}\gamma\cdot h\gamma\bigr)
&\le 
\mu^2\gamma_i\chi^2f^{-1}\Theta^{2s}\gamma_i
+C_9Q
\\&
\le 
\epsilon\gamma^2\chi^2f^{-1}\Theta^{2s}\gamma^2
+{C_{10}\epsilon^{-1}\mu^4\chi^2f^{-1}\Theta^{2s}}
+C_9Q,
\end{align*}
the second term on the right-hand side of which is further bounded as
\begin{align}
\begin{split}
C_{10,\epsilon}\mu^4\chi^2f^{-1}\Theta^{2s}
&
\leq
-C_{10,\epsilon}\mu^3\mathop{\mathrm{Im}}\bigl(\chi^2f^{-1}\Theta^{2s}(H-z)\bigr)
\\&\phantom{{}={}}{}
+\tfrac12C_{10,\epsilon}\mu^3\mathop{\mathrm{Re}}\bigl((\nabla \chi^2f^{-1}\Theta^{2s})\cdot \gamma\bigr)
\\&\phantom{{}={}}{}
+\tfrac12C_{10,\epsilon}\mu^3(\nabla \chi^2f^{-1}\Theta^{2s})\cdot (\nabla\theta)
\\&
\le 
C_{11,\epsilon}\mu^3\chi^2f^{-2}\Theta^{2s}
+\epsilon \gamma_\|\chi^2fr^{-1}\Theta^{2s}\gamma_\|
+C_{11,\epsilon}Q
\\&
\le 
2\epsilon \gamma_\|\chi^2fr^{-1}\Theta^{2s}\gamma_\|
+C_{12,\epsilon}Q
.
\end{split}
\label{eq:20091911}
\end{align}
In the third inequality of \eqref{eq:20091911} we repeated the arguments for the first and the second,
by which we can reduce the order of $f$ as much as that of $\mu$. 
After some further estimation this leads to the
 following bound of the first term of \eqref{eq:20091716}, 
\begin{align}
\begin{split}
\mathop{\mathrm{Re}}\bigl(\gamma_\| \chi^2\Theta^{2s}\gamma\cdot h\gamma\bigr)
&
\le 
-\tfrac12\gamma_j\gamma_i\chi^2f^{-1}\Theta^{2s}\gamma_i\gamma_j
+C_{13}\gamma_j\gamma_i\chi^2f^{-3}\Theta^{2s}\gamma_i\gamma_j
\\& \phantom{{}={}}{}
+2\epsilon\gamma^2\chi^2f^{-1}\Theta^{2s}\gamma^2
+2\epsilon \gamma_\|\chi^2 fr^{-1} \Theta^{2s}\gamma_\| 
\\&\phantom{{}={}}{}
-(\tfrac14-\epsilon)\gamma_i\chi^2f^{-1}g^4r^{-1}\Theta^{2s} \gamma_i
\\&\phantom{{}={}}{}
+\epsilon\chi^2f^{-1}r^{-2}g^8\Theta^{2s}
+{C_{13,\epsilon}Q}
.
\end{split}
\label{eq:200919}
\end{align}
We further bound the first and second terms of \eqref{eq:200919} as,
letting $K>0$ be large enough and using Lemma~\ref{lem:algebra0},
\begin{align*}
&-\tfrac12\gamma_j\gamma_i\chi^2f^{-1}\Theta^{2s}\gamma_i\gamma_j
+C_{13}\gamma_j\gamma_i\chi^2f^{-3}\Theta^{2s}\gamma_i\gamma_j
\\&
\le
-(\tfrac12-\epsilon )\gamma_j\gamma_i\chi^2f^{-1}\Theta^{2s}\gamma_i\gamma_j
+C_{14}\gamma_j\gamma_i\chi^2\chi(f<K)\Theta^{2s}\gamma_i\gamma_j
\\&
\le
-(\tfrac12-2\epsilon) \gamma^2\chi^2f^{-1}\Theta^{2s}\gamma^2
+C_{15,\epsilon}Q,
\end{align*}
and hence at last the first term of \eqref{eq:20091716} is bounded as 
\begin{align}
\begin{split}
\mathop{\mathrm{Re}}\bigl(\gamma_\| \chi^2\Theta^{2s}\gamma\cdot h\gamma\bigr)
&
\le 
-(\tfrac12-4\epsilon)\gamma^2\chi^2f^{-1}\Theta^{2s}\gamma^2
\\&\phantom{{}={}}{}
+2\epsilon \gamma_\|\chi^2fr^{-1} \Theta^{2s}\gamma_\|
-(\tfrac14-\epsilon)\gamma_i\chi^2f^{-1}g^4r^{-1}\Theta^{2s} \gamma_i
\\&\phantom{{}={}}{}
+\epsilon\chi^2f^{-1}r^{-2}g^8\Theta^{2s}
+{C_{16,\epsilon}Q}.
\end{split}
\label{eq:2009192}
\end{align} 

 The second and third terms of \eqref{eq:20091716} are bounded in the same manner
by substituting the second expression of Lemma~\ref{eq:200807b} 
and using the Cauchy-Schwarz inequality and Lemmas~\ref{lem:20092016}, \ref{lem:basic30} and \ref{lem:algebra0}. 
We omit the details, and directly present the resulting bounds. 
As for the second term of \eqref{eq:20091716}, we have 
\begin{align*}
&\tfrac s2\mathop{\mathrm{Re}}\bigl(\gamma_\| \chi^2r^{-1}\Theta'\Theta^{2s-1}\gamma_\|^2\bigr)
\\&
=
s\mathop{\mathrm{Re}}\bigl((H-z)^*(f^2+\lambda)^{-1} \chi^2\Theta'\Theta^{2s-1}\gamma_\|^2\bigr)
+s\mu\mathop{\mathrm{Im}}\bigl((f^2+\lambda)^{-1} \chi^2\Theta'\Theta^{2s-1}\gamma_\|^2\bigr)
\\&\phantom{{}={}}{}
-\tfrac s2\mathop{\mathrm{Re}}\bigl(\gamma^2(f^2+\lambda)^{-1} \chi^2\Theta'\Theta^{2s-1}\gamma_\|^2\bigr)
- \tfrac s4\mathop{\mathrm{Im}}\bigl(fr^{-2}(f^2+\lambda)^{-1}(g^2-\lambda) \chi^2\Theta'\Theta^{2s-1}\gamma_\|^2\bigr)
\\&\phantom{{}={}}{}
-\tfrac s4\mathop{\mathrm{Re}}\bigl(r^{-1}(f^2+\lambda)^{-1}(g^2-\lambda)^2 \chi^2\Theta'\Theta^{2s-1}\gamma_\|^2\bigr)
-s\mathop{\mathrm{Re}}\bigl((f^2+\lambda)^{-1}q_1 \chi^2\Theta'\Theta^{2s-1}\gamma_\|^2\bigr),
\end{align*}
and thus 
\begin{align}
\begin{split}
&\tfrac s2\mathop{\mathrm{Re}}\bigl(\gamma_\| \chi^2r^{-1}\Theta'\Theta^{2s-1}\gamma_\|^2\bigr)
\\&
\le 
\epsilon\gamma^2\chi^2f^{-1}\Theta^{2s}\gamma^2
+\epsilon \gamma_\|\chi^2fr^{-1}\Theta^{2s}\gamma_\|
+\epsilon\chi^2f^{-1}r^{-2}g^8\Theta^{2s}
+C_{17,\epsilon}Q.
\end{split}
\label{eq:2009193}
\end{align}
Here we note that we have discarded two non-positive operators since 
they are unnecessary for the later arguments.
As for the third term of \eqref{eq:20091716}, we have 
\begin{align*}
\tfrac14\mathop{\mathrm{Re}}\bigl(\gamma_\| \chi^2 fg^4r^{-2}\Theta^{2s}\bigr)
&
=
\tfrac12\mathop{\mathrm{Re}}\bigl((H-z)^*(f^2+\lambda)^{-1}\chi^2fg^4r^{-1}\Theta^{2s}\bigr)
\\&\phantom{{}={}}{}
-\tfrac14\mathop{\mathrm{Re}}\bigl(\gamma^2(f^2+\lambda)^{-1}\chi^2fg^4r^{-1}\Theta^{2s}\bigr)
\\&\phantom{{}={}}{}
-\tfrac18(f^2+\lambda)^{-1}(g^2-\lambda)^2\chi^2fg^4r^{-2}\Theta^{2s}
\\&\phantom{{}={}}{}
-\tfrac12(f^2+\lambda)^{-1}q_1\chi^2fg^4r^{-1}\Theta^{2s}
,
\end{align*}
and thus 
\begin{align}
\begin{split}
\tfrac14\mathop{\mathrm{Re}}\bigl(\gamma_\| \chi^2 fg^4r^{-2}\Theta^{2s}\bigr)
&
\le 
-(\tfrac14-\epsilon)\gamma_i\chi^2f^{-1}g^4r^{-1}\Theta^{2s}\gamma_i
\\&\phantom{{}={}}{}
-(\tfrac18-\epsilon)\chi^2f^{-1}r^{-2}g^8\Theta^{2s}
+C_{18,\epsilon
}Q
.
\end{split}
\label{eq:20091820}
\end{align}
Hence by \eqref{eq:20091716}, \eqref{eq:2009192},
\eqref{eq:2009193} and \eqref{eq:20091820} we obtain 
\begin{align}
\begin{split}
\mathop{\mathrm{Im}}(P(H-z))
&
\le 
-(\tfrac12-5\epsilon)\gamma^2\chi^2f^{-1}\Theta^{2s}\gamma^2
-(\tfrac12-2\epsilon)\gamma_i\chi^2f^{-1}g^4r^{-1}\Theta^{2s} \gamma_i
\\&\phantom{{}={}}{}
-\tfrac12\gamma_\| \chi^2fg^2r^{-2}\Theta^{2s}\gamma_\|
+(\tfrac s2+8\epsilon)\gamma_\| \chi^2 fr^{-1}\Theta^{2s}\gamma_\|
\\&\phantom{{}={}}{}
-(\tfrac18-5\epsilon)\chi^2f^{-1}r^{-2}g^8\Theta^{2s}
+{C_{19,\epsilon}Q}
.
\end{split}
\label{eq:200919249}
\end{align}

\smallskip
\noindent
\emph{Step IV}.
Finally we produce a good negative quantity by combining 
the first, second and fifth terms of \eqref{eq:200919249}. 
For that we compute 
\begin{align*}
&
\gamma_\| \chi^2 f^3r^{-2}\Theta^{2s}\gamma_\|
\\&
=
4
\Bigl(H-z^*-\mathrm i\mu-\tfrac12\gamma^2-\tfrac{\mathrm i}4fr^{-2}(g^2-\lambda)-\tfrac14r^{-1}(g^2-\lambda)^2-q_1\Bigr)
\\&\phantom{{}={}}{}
\cdot\chi^2f^3(f^2+\lambda)^{-2}\Theta^{2s}
\Bigl(H-z+\mathrm i\mu-\tfrac12\gamma^2+\tfrac{\mathrm i}4fr^{-2}(g^2-\lambda)-\tfrac14r^{-1}(g^2-\lambda)^2-q_1\Bigr)
\\&
\le 
(1+\epsilon)\Bigl(\gamma^2\chi^2f^{-1}\Theta^{2s}\gamma^2
+\gamma_i\chi^2f^{-1}g^4r^{-1}\Theta^{2s}\gamma_i
+\tfrac14\chi^2f^{-1}r^{-2}g^8\Theta^{2s}\Bigr)
\\&\phantom{{}={}}{}
+C_{20}\mu^2\chi^2f^{-1}\Theta^{2s}
+C_{20,\epsilon}Q
.
\end{align*}
The above second term with the factor $\mu^2$ can be bounded similarly to \eqref{eq:20091911},
and then we obtain 
\begin{align}
\begin{split}
&
-\gamma^2\chi^2f^{-1}\Theta^{2s}\gamma^2
-\gamma_i\chi^2f^{-1}g^4r^{-1}\Theta^{2s}\gamma_i
-\tfrac14\chi^2f^{-1}r^{-2}g^8\Theta^{2s}
\\&
\le 
-(1-\epsilon)\gamma_\| \chi^2 f^3r^{-2}\Theta^{2s}\gamma_\|
+\epsilon\gamma_\|\chi^2fr^{-1}\Theta^{2s}\gamma_\|
+{C_{21}\mu f^{-2}\Theta^{2s}}
+C_{21,\epsilon}Q
.
\end{split}
\label{eq:200911118}
\end{align}
{We remark that the second term on the right-hand
 side of the bound of the lemma shows up here.}
We combine \eqref{eq:200919249} and \eqref{eq:200911118}, and hence obtain 
\begin{align*}
\begin{split}
\mathop{\mathrm{Im}}(P(H-z))
&
\le 
-\tfrac12(1-40\epsilon)(1-\epsilon)\gamma_\| \chi^2 f^3r^{-2}\Theta^{2s}\gamma_\|
-\tfrac12\gamma_\| \chi^2fg^2r^{-2}\Theta^{2s}\gamma_\|
\\&\phantom{{}={}}{}
+\tfrac12(s+17\epsilon)\gamma_\| \chi^2 fr^{-1}\Theta^{2s}\gamma_\|
+{C_{22}\mu f^{-2}\Theta^{2s}}
+C_{22,\epsilon}Q
\\&
\le 
-\tfrac12(2-s-99\epsilon)\gamma_\| \chi^2fr^{-1}\Theta^{2s}\gamma_\|
+{C_{22}\mu f^{-2}\Theta^{2s}}
+C_{22,\epsilon}Q
.
\end{split}
\end{align*}
Now by letting $\epsilon\in (0,1)$ be small enough the lemma follows. 
\end{proof}
\begin{remark}\label{rem:20:1030}
By an argument similar to \eqref{eq:20091911} we can replace 
the second term in Lemma \ref{lem:200919} by $Cf^{-3}\Theta^{2s}$.
See Remarks~\ref{rem:20103018} \ref{item:2010301809} and 
\ref{remarks:commutator-estimates} \ref{item:20103015} for an application of this fact.
\end{remark}

Next we bound the quantity \eqref{eq:20090912} from below, using again
 the form $Q_s$ from \eqref{eq:bookk}. 
\begin{lemma}\label{lem:200919b}
Let $I\subseteq \mathbb R$ be a compact interval, 
$\kappa>3$ large enough, and $\kappa'\in (0,1/2)$. 
For any $s>0$ and $\epsilon\in (0,1)$ 
there exist $C>0$ such that 
uniformly in $\nu\in\mathbb N_0$ and $z=\lambda\pm \i\mu\in I_\pm$,
as a form on $\vH^2_{-\infty}$,
\begin{align*}
\pm\mathop{\mathrm{Im}}(P(H-z))
\ge 
-\epsilon\gamma_\|\chi^2 f^{-1}\Theta^{2s}\gamma_\|
-{C\mu f^{-2}\Theta^{2s}}-CQ,
\end{align*} 
where 
\begin{align*}
Q
=
f^{-3-2\delta}\Theta^{2s}
+(H-z)^*f^5\Theta^{2s}(H-z) +Q_s.
\end{align*} 
\end{lemma}
\begin{proof}
This is due to computations similar to but shorter than those in the proof of
Lemma~\ref{lem:200919}.
We prove the assertion only for the upper sign since the lower one follows by the same manner. 
{Parallelly} to the proof of Lemma \ref{lem:200919} a term bounded by $Q$ 
may be called an \emph{admissible error term} (note that the present
$Q$ is slightly different).
Then by \eqref{eq:200820}, Lemmas~\ref{lem:20092016} and \ref{lem:basic30} and the Cauchy-Schwarz inequality 
\begin{align}
\begin{split}
\mathop{\mathrm{Im}}(P(H-z))
&=
\mathop{\mathrm{Im}}\!\parb{\gamma_\|^2\chi^2\Theta^{2s}(H-z)}
+2\mathop{\mathrm{Re}}\!\parb{\gamma_\|r(\nabla f)\cdot (\nabla \chi^2\Theta^{2s})(H-z)}
\\&\ge 
-\epsilon \gamma_\|^2\chi^2 f^{-1}(2r)^{-2}\Theta^{2s}\gamma_\|^2
-\epsilon \gamma_\|\chi^2f^{-1}\Theta^{2s}\gamma_\|
-C_{1,\epsilon}Q
.
\end{split}
\label{eq:2009191314}
\end{align}
Here we note the following kind of converse of \eqref{eq:200911118}, 
\begin{align}
\begin{split}&
\gamma^2\chi^2f^{-1}\Theta^{2s}\gamma^2
+\gamma_i\chi^2f^{-1}g^4r^{-1}\Theta^{2s}\gamma_i
+\tfrac14\chi^2f^{-1}r^{-2}g^8\Theta^{2s}
\\&
\le 
C_2\gamma_\| \chi^2 f^{-1}\Theta^{2s}\gamma_\|
+{C_{2}\mu f^{-2}\Theta^{2s}}
+C_{2}Q,
\end{split}
\label{eq:20091913}
\end{align}
 which can be proved similarly to \eqref{eq:200911118}. 
Then the first term on the right-hand side of \eqref{eq:2009191314}
may be 
estimated, using in the first line of Lemma \ref{lem:algebra0}, as 
\begin{align*}
-\epsilon \gamma_\|^2\chi^2 f^{-1}(2r)^{-2}\Theta^{2s}\gamma_\|^2
&\ge 
-2\epsilon \gamma^2\chi^2f^{-1}\Theta^{2s}\gamma^2-C_3Q
\\&
\ge 
-C_4\epsilon \gamma_\|\chi^2f^{-1}\Theta^{2s}\gamma_\|
-{C_4\mu f^{-2}\Theta^{2s}}
-C_4Q.
\end{align*} 
Hence we obtain the asserted bound. 
\end{proof}

\subsection{
 Proposition~\ref{prop:phase-space-radiation} finalized}\label{subsec:propPr} 
The essential parts of our proof of
Proposition~\ref{prop:phase-space-radiation} are already done in Lemmas~\ref{lem:200912}, \ref{lem:200913}, 
\ref{lem:200919} and \ref{lem:200919b}.
 Let $I\subseteq\mathbb R$ be a compact interval, and let $s\in(0,1+\delta)$. 
We {start with} a classically allowed region. 
Let $\kappa>3$ be sufficiently large, and $\kappa'\in (0,1/2)$. 
By Lemmas~\ref{lem:algebra0}, \ref{lem:200919} and \ref{lem:200919b} along with \eqref{eq:20091913} 
it follows that uniformly in $\nu\in\mathbb N_0$ and $z=\lambda\pm
\i\mu\in I_\pm$
\begin{align*}
&\gamma_j\gamma_i\chi^2 f^{-1}\Theta^{2s}\gamma_i\gamma_j
+\chi^2 f^3\Theta^{2s}\tilde\gamma^8
+\gamma_\|\chi^2 f^{-1}\Theta^{2s}\gamma_\|
\\&\le 
{C_1\mu f^{-2}\Theta^{2s}}
+C_1f^{-3-2\delta}\Theta^{2s}
+C_1(H-z)^*f^5\Theta^{2s}(H-z) +C_1Q_s.
\end{align*} 
Take its expectation in the state $\phi=R(z)\psi$ with $\psi\in C^\infty_{\mathrm c}(\mathbb R^d)$. 
Then, due to 
the Cauchy-Schwarz inequality, 
Theorem~\ref{thm:200909} and Lemmas~\ref{lem:200912} and \ref{lem:200913}, 
uniformly in $\nu\in\mathbb N_0$ and $z\in I_\pm$
\begin{align*}
&
\|\chi \Theta^{s}\gamma_i\gamma_j\phi\|_{L^2_{-1/2}}^2
+\|\chi f\Theta^{s/2}\tilde\gamma^2\phi\|_{\mathcal B^*}^2
+\|\chi\Theta^{s}\gamma_\|\phi\|_{L^2_{-1/2}}^2
\\&\le 
\|\chi \Theta^{s}\gamma_i\gamma_j\phi\|_{L^2_{-1/2}}^2
+\|\chi \phi\|_{\mathcal B^*}^2
+\|\chi f^2\Theta^{s}\tilde\gamma^4\phi\|_{L^2_{-1/2}}^2
+\|\chi\Theta^{s}\gamma_\|\phi\|_{L^2_{-1/2}}^2
\\&\le 
{C_{\nu}\mu\|\psi\|_{\mathcal B}^2}
+C_2\|\psi\|_{\mathcal B}^2
+C_2\|f^{s+3/2}\psi\|_{L^2_{1/2}}^2
+C_1\|f^{s+2}\psi\|_{L^2_{1/2}}^2
\\&
\le 
{C_{\nu}\mu\|\psi\|_{\mathcal B}^2}
+C_3\|f^{s+2}\psi\|_{L^2_{1/2}}^2,
\end{align*} 
 {where the constant $C_\nu$ depends on $\nu\in\mathbb N_0$ defining $\Theta$}.
By Lemmas~\ref{lem:200912} and \ref{lem:200913} again, 
the same bounds hold with $\chi$ replaced by $\sqrt{1-\chi^2}$, 
hence we finally conclude the uniform bounds 
\begin{align}\label{eq:Fbndb}
 \begin{split}
&\| \Theta^{s}\gamma_i\gamma_j\phi\|_{L^2_{-1/2}}^2
+\| f\Theta^{s/2}\tilde\gamma^2\phi\|_{\mathcal B^*}^2
+\|\Theta^{s}\gamma_\|\phi\|_{L^2_{-1/2}}^2
\\&
\le 
{C_{\nu}\mu\|\psi\|_{\mathcal B}^2}
+C_4\|f^{s+2}\psi\|_{L^2_{1/2}}^2.
 \end{split}
\end{align}
Now let $\mu\to 0$, for example by invoking Fatou's lemma and a
diagonal sequence argument, and then $\nu\to\infty$ by invoking the monotone convergence
 theorem. Whence Proposition~\ref{prop:phase-space-radiation}
 follows.
\begin{remarks}\label{remarks:commutator-estimates}
\begin{enumerate}[1)]
\item
We can bound $\|f^{s+2}\tilde\gamma^4\phi\|_{L^2_{-1/2}}$ 
if a stronger weight is used on the right-hand side of \eqref{eq:p1}.
For a classically allowed region this can be seen from the above proof 
without any additional weight,
while for a classically forbidden region we in fact need an additional weight.
\item \label{item:20103015}
As observed in Remark~\ref{rem:20:1030} we can bound 
$\mu f^{-2}\Theta^{2s}$ by $f^{-3}\Theta^{2s}$.
Then for $s\in (0,1)$ the bounds in the above proof become uniform in $\nu\in\mathbb N_0$,
i.e.\ we may let $C_\nu=0$. 
Letting $\nu\to\infty$ we obtain this way 
 uniform bounds in the complex spectral parameter,
 cf. Remarks~\ref{rem:20103018} \ref{item:2010301809}. We can not
 prove similar uniform bounds for $s\in [1, 1+\delta)$.
\item \label{item:bProof}
The bounds \eqref{eq:p3} follow readily by
 integration by parts and using the Cauchy-Schwarz inequality as 
mentioned in {Remarks~\ref{rem:20103018} \ref{item:2010301810}, however they 
should be applied along with Theorem \ref{thm:200909} and \eqref{eq:Fbndb}
\emph{before} taking the limits $\mu\to 0$ and $\nu\to\infty$}.
\end{enumerate}
\end{remarks}

\subsection*{Acknowledgements} 
{T.A. is supported by JSPS KAKENHI grant nr.\ 17K05319.
K. Ito is supported by JSPS KAKENHI grant no.\ 17K05325.
E.S. is supported by the Research Institute for Mathematical Sciences, a Joint
Usage/Research Center located in Kyoto University, and by DFF grant
no.\ 4181-00042. K. Ito and E.S. are supported by the Swedish Research
Council grant no.\ 2016-06596 (residing at Institut Mittag-Leffler in Djursholm, Sweden, during the Spring semester of 2019).}

\end{document}